%% file: paper.tex
\documentclass[letterpaper,twocolumn,screen,10pt]{article}
\usepackage{usenix}

\input{includes/packages}
\input{includes/settings}

\if\showcomments1
    \usepackage[colorinlistoftodos,textsize=tiny]{todonotes}
\else
    \usepackage[disable]{todonotes}
\fi

\begin{document}

\title{\titleinfo}

\if\shownames1
    \author{\authorinfo \\ \institutioninfo}
\else
    \author{\submissioninfo}
\fi

\if\showcomments1
    \onecolumn
    \setcounter{page}{0}
    \listoftodos{}
    \clearpage
    \twocolumn
    \setcounter{page}{1}
\fi

\if\showpagenumbers0
	\pagestyle{empty}
\fi

\maketitle

\input{sections/abstract}

\input{sections/introduction}
\input{sections/background}
\input{sections/motivation}
\input{sections/design}
\input{sections/implementation}
\input{sections/evaluation}
\input{sections/related-work}
\input{sections/conclusion}

\label{lastpage}

\if\showacks1
    \input{sections/acknowledgements}
\fi

{\small
\bibliographystyle{plain}
\bibliography{bibs/paper}}

\input{sections/appendix}

\label{totalpage}

\end{document}

%% file: includes/packages.tex
\usepackage{xspace} 
\usepackage{enumitem}
\usepackage{color}
\usepackage{colortbl}
\usepackage{xcolor}
\usepackage{pifont}

\usepackage{epstopdf}
\usepackage{graphicx}
\usepackage[labelfont=bf, textfont=bf]{caption}
\usepackage{subcaption}
\usepackage{pgfplots}
\usepackage{pgfplotstable}

\usepackage{siunitx}
\usepackage{ifthen}

\usepackage{listings}
\usepackage[utf8]{inputenc}

\usepackage{booktabs}  
\usepackage{tabu}
\usepackage{array}
\usepackage{multirow}
\usepackage{diagbox}
\usepackage{tablefootnote}

\usepackage{balance}
\usepackage{microtype}

\usepackage[most]{tcolorbox}

\usepackage[english]{babel}
\usepackage{blindtext}

\usepackage[capitalise]{cleveref}

\usepackage{xurl}

%% file: includes/settings.tex
\def\setuppreprint{1}
\def\setupcameready{0}

\newcommand{\name}{\textsc{Tegra}\xspace}

\newcommand{\titleinfo}{\name{}: A Flexible \& Scalable NextGen Mobile Core\vspace{-5pt}}

\newcommand{\institutioninfo}{Purdue~~~$^\dagger$Intel~~~$^\ddagger$Michigan, Ann Arbor~~~$^\diamond$UT~Arlington~~~$^\star$Princeton~~~$^\circ$Unaffiliated}
\newcommand{\authorinfo}{\em Bilal Saleem, Omar Basit, Jiayi Meng$^\diamond$, Iftekhar Alam$^\dagger$, Ajay Thakur$^\circ$,\\\em Christian Maciocco$^\dagger$, Muhammad Shahbaz$^\ddagger$, Y. Charlie Hu, Larry Peterson$^\star$}
\newcommand{\submissioninfo}{\em Submission \#1 \\ {\small \pageref{lastpage} Pages Body, \pageref{totalpage} Pages Total}}

\if\setupcameready1
	\def\shownames{1} 										
	\def\showpagenumbers{0}                                 
    \def\showcomments{0} 									
    \def\showacks{1} 										
\else
	\def\shownames{1}										
	\def\showpagenumbers{0}                                 
    \if\setuppreprint1										
        \def\showcomments{0}
    \else
        \def\showcomments{1}
    \fi
    \def\showacks{0}										
\fi

\newcommand{\ie}{i.e.}
\newcommand{\eg}{e.g.}
\newcommand{\ea}{et al.}

\setlist[itemize]{topsep=4pt, itemsep=4pt, parsep=1.5pt}

\newcommand{\cmark}{\ding{51}}
\newcommand{\xmark}{\ding{55}}

\newcommand{\ballnumber}[1]{\tikz[baseline=(myanchor.base)] \node[circle,fill=red,inner sep=1pt] (myanchor) {\color{-.}\bfseries\footnotesize #1};}

\newcommand{\boxtext}[1]{\tikz[baseline=(myanchor.base)] \node[rectangle,fill=black,inner sep=1pt] (myanchor) {\color{-.}\bfseries\footnotesize #1};}

%% file: sections/abstract.tex
\begin{abstract}
To support emerging mobile use cases (\eg, AR/VR, autonomous driving, and massive IoT), next-generation mobile cores for 5G and 6G are being re-architected as service-based architectures (SBAs) running on both private and public clouds.
However, current performance optimization strategies for scaling these cores still revert to traditional NFV-based techniques, such as consolidating functions into rigid, monolithic deployments on dedicated servers.
This raises a critical question: Is there an inherent tradeoff between {\em flexibility} and {\em scalability} in an SBA-based mobile core, where improving performance (and resiliency) inevitably comes at the cost of one or the other?

To explore this question, we introduce resilient SBA microservices design patterns and state-management strategies, and propose \name{}---a high-performance, flexible, and scalable SBA-based mobile core. 
By leveraging the mobile core's unique position in the end-to-end internet ecosystem (\ie, at the last-mile edge), \name{} optimizes performance without compromising adaptability. 
Our evaluation demonstrates that \name{} achieves significantly lower latencies, processing requests 20$\times$, 11$\times$, and 1.75$\times$ faster than traditional SBA core implementations---free5GC, Open5GS, and Aether, respectively---all while matching the performance of state-of-the-art cores (\eg, CoreKube) while retaining flexibility.
Furthermore, it reduces the complexity of deploying new features, requiring orders of magnitude fewer lines of code (LoCs) compared to existing cores.
\end{abstract}

%% file: sections/introduction.tex
\vspace{5pt}
\section{Introduction}
\label{sec:intro}

As we move into an increasingly connected and inter-networked society, humans are no longer the primary consumers of data~\cite{ahmad2020low}.
With trends like massive Internet-of-Things (IoT), self-driving cars, and augmented and virtual reality (AR/VR) taking a stronghold~\cite{ahmad2020low}, the human-to-data (or human-to-human) interactions on the Internet are being vastly subsumed by the device-to-data (and device-to-device) interactions.
The mobile (or cellular) core is at the heart of this transition and will define how the future will take shape~\cite{ahmad2020low, peterson2019democratizing}.
In response, the cellular network community is already breaking away from the rigid and purpose-built mobile-core architectures of 4G and earlier, \eg, fixed-function hardware~\cite{basta2013virtual} or Network Function (NF) virtualization (NFV)-based appliances~\cite{nguyen2017sdn}, to better accommodate the diverse and evolving needs of emerging classes of users and workloads (\eg, eMBB, URLLC, and massive IoT)~\cite{khan2022urllc}.
However, this shift is putting considerable strain on the control plane, particularly in scenarios where thousands of IoT devices repeatedly connect and disconnect while exchanging minimal data.

\begin{figure}[t]
\centering
\includegraphics[width=\linewidth]{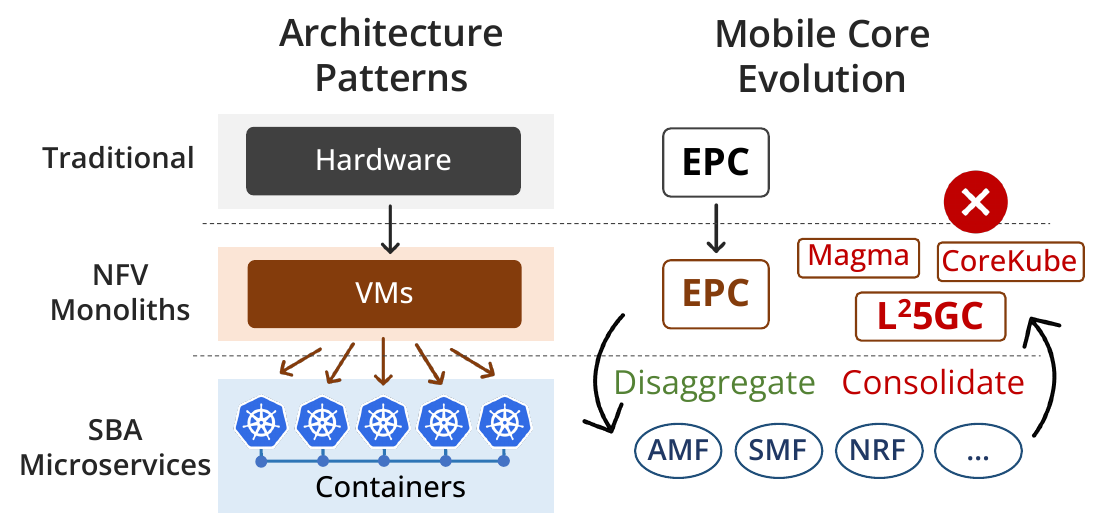}
\caption{To gain flexibility and agility, mobile cores have evolved from traditional proprietary hardware to custom-built monolithic NFV-based VMs, to now a fine-grained collection of microservices (\eg, AMF and SMF). 
Yet, for performance reasons, current proposals (\eg, L$^2$5GC, Magma, and CoreKube) revert to consolidating services using past NFV-based approaches.}
\label{fig:nfv-sba}
\vspace{-15pt}
\end{figure}

Starting with 5G and onwards, mobile core providers are embracing a {\em service-based architecture (SBA)}~\cite{brown2017service, 5gsysapp}, where the traditional monolithic processing stack (\eg, NFV-based EPC~\cite{nguyen2017sdn}) is decomposed into distinct components (\eg, AMF and SMF)~\cite{brown2017service}\footnote{We describe these 5G core functions in \Cref{sec:app-5gc-funcs}.} that operate as disaggregated services, in particular, microservices~\cite{nokia-cloudnative} (\Cref{fig:nfv-sba}).
These microservices communicate with each other using a service-based interface (SBI) and standard APIs (\eg, HTTPS/REST), which enable telco operators to easily add, manage, and upgrade new features (via CI/CD~\cite{rajavaram2019automation})---incorporating them as flexible, pluggable services~\cite{disaggregation-5g-services}, either homegrown or from other vendors (\eg, AMF from Nokia~\cite{nokia-AMF} and SMF from Cisco~\cite{cisco-SMF}).
Moreover, these services can run alongside other tenants' workloads across on-premise, edge, or public clouds~\cite{nokia-cloudnative}, significantly reducing telcos' CAPEX/OPEX while improving resource efficiency.

Despite its flexible and scalable design, recent proposals~\cite{jain2022l25gc, mohammadkhan2020cleang, buyakar2019prototyping, liao2020evaluating, vittal2021zero, ferguson2023corekube} aimed at improving the end-to-end performance (and resiliency) of SBA-based mobile cores have resorted to past methods and techniques used in optimizing NFV-based monolithic core designs (\eg, consolidating and aggregating logic and compute onto dedicated servers)~\cite{madhura2020dynamic}, \Cref{fig:nfv-sba}. 
For example, state-of-the-art proposals either {\em (a)} bundle the 5G core services (\eg, AMF and SMF) onto a single server or VM by replacing the HTTPS/REST-based SBI with a shared-memory interface to reduce inter-service delays (\eg, L$^2$5GC~\cite{jain2022l25gc}) or {\em (b)} code them into a single (monolithic) binary running inside a stateless container alongside persistent storage (\eg, Magma~\cite{magma} and CoreKube~\cite{ferguson2023corekube}). 
Doing so yields better end-to-end latency but at the cost of flexibility---the very reason why the 5G community adopted SBA in the first place, \ie, to break vendor lock-in~\cite{vendorLockIn} and enable seamless creation, integration, and interoperability of new services from different vendors (\eg, Nokia, Cisco, Ericsson, or Intel) ~\cite{disaggregation-5g-services, nokia-AMF, cisco-SMF, EricssonCore, IntelCore}.

There are two reasons why, we believe, the cellular community is conflicted in its approach to making the modern SBA-based cores {\em fast} (reverting to consolidation) and {\em flexible} (opting for disaggregation).
First, treating microservices of an SBA-based core the same as NFV-based elements of a traditional core (\ie, Network Functions or NFs) obscures the distinction between the two---resulting in proposals that apply the same optimization techniques they used for monoliths to microservices~\cite{jain2022l25gc, ferguson2023corekube}.
Second, more importantly, overlooking key insights into the workings of these SBA-based cores and their unique position in the network (\ie, servicing bidirectional upstream and downstream events at the edge)~\cite{aether, free5gc, oai} further hampers us from accurately identifying and addressing the real sources of performance bottlenecks.

To bridge this gap, we present \name{}, a high-performance, scalable, and resilient SBA-based 5G core that carefully balances efficiency and flexibility.
\name{} introduces scalable and resilient microservices design patterns and state-management strategies that enable fine-grained autoscaling, load balancing, and fault tolerance within the SBA-based mobile core architecture. 
Unlike fully stateless designs~\cite{ferguson2023corekube} that require frequent interactions with external storage, \name{} adopts a soft-state approach, maintaining per-user-equipment (UE) context within individual service instances while asynchronously persisting updates to storage. 
This method significantly reduces the overhead of continuous storage access, enabling low-latency message processing while ensuring service continuity during failures.

To enhance fault tolerance, \name{} ensures that new instances can swiftly recover the latest committed state when failures occur, preventing service disruptions while keeping storage overhead minimal. 
Additionally, it implements UE-based sticky load balancing, ensuring that all upstream and downstream control messages for a given UE are consistently handled by the same service instance. 
By localizing UE-related state, this approach eliminates inconsistencies, reduces unnecessary storage lookups, and optimizes end-to-end performance---delivering high efficiency without relying on NFV-like solutions (\eg, consolidating services~\cite{ferguson2023corekube, jain2022l25gc}).
By integrating these design principles, \name{} provides a robust solution that enhances fault tolerance, improves resource efficiency, and outperforms existing SBA-based mobile cores such as Aether~\cite{aether-deploy}, free5GC~\cite{free5gc}, Open5GS~\cite{open5gs}, CoreKube~\cite{ferguson2023corekube}, and Magma~\cite{magma}. 
At the same time, it fully retains the modular flexibility of SBA, enabling seamless integration with 3rd-party services and cloud-native deployments.

Our evaluation (\Cref{sec:evaluation}) show that \name{} achieves significantly lower latencies, processing requests an order of magnitude faster than traditional SBA core implementations (\eg, free5GC~\cite{free5gc}, Open5GS~\cite{open5gs}, and Aether~\cite{aether-deploy}, respectively) all while matching the performance of state-of-the-art cores (\eg, CoreKube~\cite{ferguson2023corekube}) while retaining flexibility.
Furthermore, it reduces the complexity of deploying new features, requiring orders of magnitude fewer lines of code (LoCs) compared to existing cores.

\begin{tcolorbox}[
    colback=blue!5!white,
    colframe=black,
    title={},
    fonttitle=\bfseries,
    fontupper=\small, 
    sharp corners,
    boxrule=0.4mm,
    coltitle=black,
    enhanced jigsaw,
    drop shadow={black!50!white}
]
\textbf{Disruptive Insight:} \textit{Classical cloud-native services---such as web or storage backends---typically run at the terminal ends of the Internet (\ie, in public clouds or enterprise datacenters), handling traffic initiated by clients. 
In contrast, the mobile core operates at an intermediary point in the network and must handle asynchronous control-plane events from both the user equipment (UE) side and the Internet (\S\ref{sec:motivation}). 
This distinct position calls for a rethinking of how core cloud-native techniques like autoscaling, load balancing, and fault tolerance are designed and implemented. 
\name{} addresses this need by adapting these mechanisms to the demands of emerging mobile core workloads, delivering high performance and flexibility without reverting to NFV-style consolidation (\S\ref{sec:design}).}
\end{tcolorbox}

\definecolor{bostonuniversityred}{rgb}{0.8, 0.0, 0.0}
\begin{table}[t]
\footnotesize
\centering
\begin{tabular}{c|ccccc|c}
  \toprule
  \textbf{5G Core} & 
  \rotatebox{90}{\textbf{Flexible}} & 
  \rotatebox{90}{\textbf{Scalable}} & 
  \rotatebox{90}{\textbf{Performant}} & 
  \rotatebox{90}{\textbf{Resilient}} & 
  \rotatebox{90}{\textbf{3GPP}} & 
  \rotatebox{0}{\textbf{Is Operational?}} \\
  \midrule
  Aether~\cite{aether-deploy}   & \textcolor{teal}{\cmark} & \textcolor{bostonuniversityred}{\xmark}  & \textcolor{bostonuniversityred}{\xmark} & \textcolor{bostonuniversityred}{\xmark} & \textcolor{teal}{\cmark} & \textcolor{teal}{\cmark} \\
  free5GC~\cite{free5gc}  & \textcolor{teal}{\cmark} & \textcolor{bostonuniversityred}{\xmark}  & \textcolor{bostonuniversityred}{\xmark} & \textcolor{bostonuniversityred}{\xmark} & \textcolor{teal}{\cmark} & \textcolor{teal}{\cmark} \\
  Open5GS~\cite{open5gs} & \textcolor{teal}{\cmark}  & \textcolor{bostonuniversityred}{\xmark}  & \textcolor{bostonuniversityred}{\xmark} & \textcolor{bostonuniversityred}{\xmark} & \textcolor{teal}{\cmark} & \textcolor{teal}{\cmark} \\
  L$^2$5GC~\cite{jain2022l25gc} & \textcolor{bostonuniversityred}{\xmark}  & \textcolor{teal}{\cmark} & \textcolor{teal}{\cmark} & \textcolor{bostonuniversityred}{\xmark} & \textcolor{teal}{\cmark} & \textcolor{teal}{\cmark} \\
  CoreKube~\cite{ferguson2023corekube} & \textcolor{bostonuniversityred}{\xmark}  & \textcolor{teal}{\cmark} & \textcolor{teal}{\cmark} & \textcolor{teal}{\cmark} & \textcolor{bostonuniversityred}{\xmark} & \textcolor{bostonuniversityred}{\xmark}\tablefootnote{CoreKube~\cite{ferguson2023corekube} is incomplete, supporting only upstream messages via SCTP (\Cref{fig:cmp-trd-sota-vs-proposed}b) and lacking support for downstream traffic over PFCP---a necessary feature for a core to be deployable in operation.
  } \\
  Magma~\cite{magma} & \textcolor{bostonuniversityred}{\xmark} & \textcolor{teal}{\cmark} & \textcolor{bostonuniversityred}{\xmark} & \textcolor{teal}{\cmark} & \textcolor{bostonuniversityred}{\xmark} & \textcolor{teal}{\cmark} \\
  \midrule
  \textbf{\name{}} & \textcolor{teal}{\cmark} & \textcolor{teal}{\cmark} & \textcolor{teal}{\cmark} & \textcolor{teal}{\cmark} & \textcolor{teal}{\cmark} & \textcolor{teal}{\cmark} \\
  \bottomrule
\end{tabular}
\vspace{-5pt}
\caption{\name{} vs. Existing 5G Cores: Unlike other solutions, \name{} excels in all key areas---flexibility, scalability, performance, and resilience---while remaining fully 3GPP-compliant and operational.}
\label{tab:core-comparison}
\vspace{-20pt}
\end{table}

%% file: sections/background.tex
\vspace{-10pt}
\section{Background} 
\label{sec:background}

\begin{figure*}[t]
\centering
\includegraphics[width=0.95\linewidth]{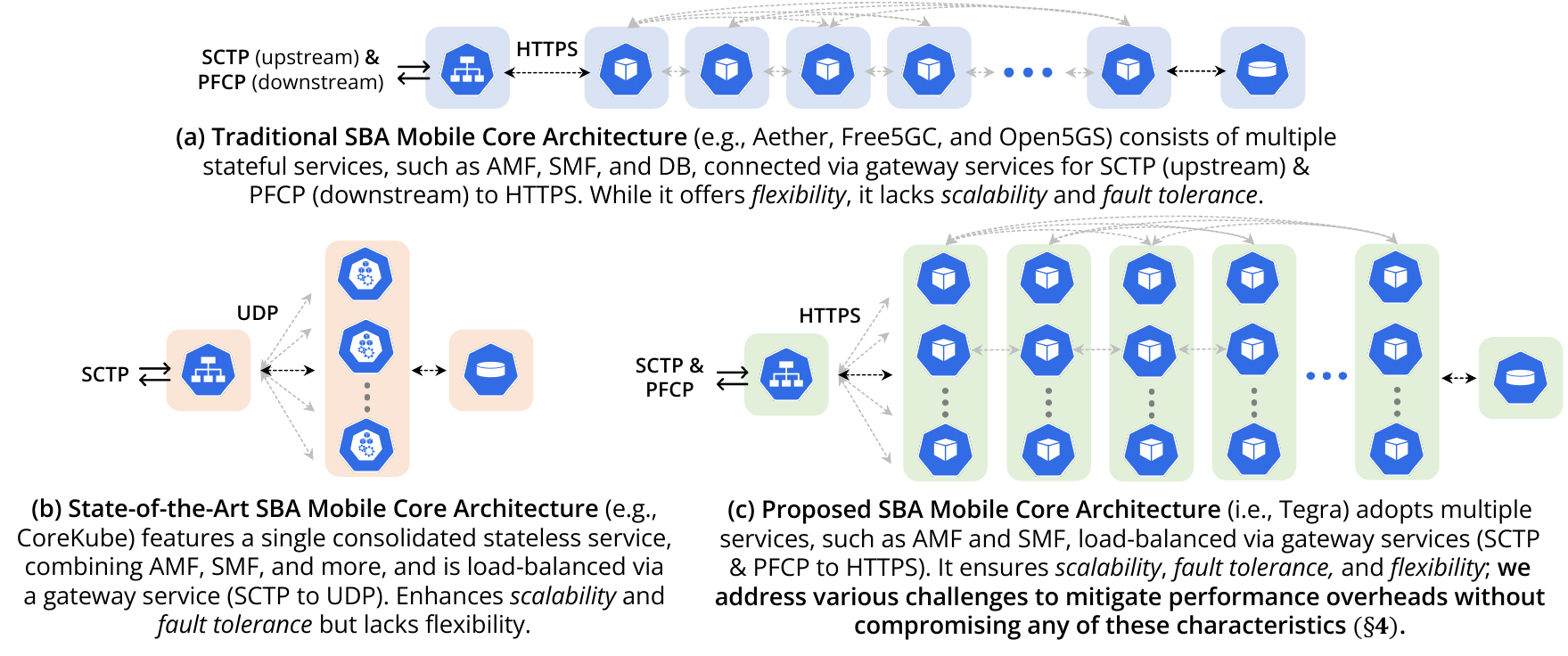}
\vspace{-8pt}
\caption{Comparing (a) traditional, (b) state-of-the-art, and (c) our proposed SBA mobile core architectures.}
\vspace{-15pt}
\label{fig:cmp-trd-sota-vs-proposed}
\end{figure*}

\vspace{-3pt}
We start with a background on the evolution of the cellular mobile core (\S\ref{ssec:trad-nfv-sba}) and the adoption of service-based architecture (SBA) in 5G (\S\ref{ssec:sba-5g}).

\subsection{Traditional vs. NFV vs. SBA Core}
\label{ssec:trad-nfv-sba}

At the time, when Software-Defined Networking (SDN)~\cite{feamster2014road} was upending the way conventional networks (\eg, campuses~\cite{lei2014swan,ghaffarinejad2014load}, enterprises~\cite{bailey2016faucet, shirmarz2020performance}, or data centers~\cite{ali2018load, shirmarz2020performance}) were controlled and managed, Network Function Virtualization (NFV)~\cite{hawilo2014nfv, nguyen2017sdn} was redefining the way cellular components were implemented and deployed.

The {\em traditional} fixed-function, proprietary hardware appliances (\eg, EPC, MME, and PACKET)~\cite{amogh2017cloud, shah2020turboepc, open5gs} were replaced with purpose-built, NFV-based appliances (deployed as virtual machines or VMs) in 4G, as shown in \Cref{fig:nfv-sba}. 
Along with compute virtualization, NFV supported high-speed network interfaces (\eg, DPDK\cite{dpdk} and SR-IOV technologies), allowing VMs to process incoming traffic with predictable latencies and throughput while coexisting on the same underlying (commodity) hardware (\eg, Intel and AMD servers).

However, with the arrival of 5G (and mmWave technology), it became challenging for telco operators to adapt their cellular networks to the continuously evolving and context-sensitive use cases (\eg, eMBB, URLLC, and massive IoT~\cite{khan2022urllc}).
The telco providers needed {\em (1)} a better and more efficient way of utilizing their infrastructure resources (VMs were too coarse-grained and bloated), {\em (2)} to adapt their environment to the changing workload demands (NFV-based appliances were too specialized and vendor-specific, resulting in longer delays for the new features to arrive), and {\em (3)} the ability to orchestrate and deploy their network elements, quickly and automatically.
Hence, starting with 3GPP release 15~\cite{3gpptr29.501}, the cellular community is shifting towards SBA-based microservices to mitigate many of the issues with NFV-based designs and to handle emerging (and future) use cases flexibly and cost-effectively. 

\subsection{Service-Based Architecture (SBA) in 5G}
\label{ssec:sba-5g}
Service-based architecture (SBA) is an emerging approach in which a single application (\eg, mobile core) is decomposed into several connected yet independently executable microservices~\cite{nokia-cloudnative, brown2017service, hirai2020automated}, deployed on a cluster of commercial off-the-shelf (COTS) servers.
Each microservice operates in isolation with a local pool of resources (\eg, compute, storage, and networking) and interacts with other services using service-based interfaces (SBIs)~\cite{5gsysapp, cao2022research, zhang2018performance} and standard APIs (\eg, HTTPS/REST~\cite{jain2022l25gc, 5gsysapp, buyakar2019prototyping}). 
SBA provides many advantages over traditional monolithic 3GPP design patterns~\cite{nfvvsclound, de2019monolithic}: {\em (i)} flexibility and agility (ease of modifying, managing, and updating services using CI/CD\footnote{CI/CD aims to realize continuous software integration,
delivery, and deployment via four phases: source, build, test, and
deploy.})~\cite{throner2021advanced,sayfan2019hands}, {\em (ii)} scalability and resource-efficiency (dynamically increasing or decreasing service instances with changing load)~\cite{coulson2020adaptive,hasselbring2017microservice}, {\em (iii)} interoperability (accessing other services via standard interfaces)~\cite{yuan2019architecture,jarwar2017exploiting,de2018pacificclouds}, {\em (iv)} security and resilience against failures (spinning up instances from the last saved state)~\cite{espinoza2022back}, and {\em (v)} vendor independence (composing services from varying vendors and developers)~\cite{mateus2021security,disaggregation-5g-services}.

Therefore, with 5G and onwards, the cellular community is redesigning the mobile core as an SBA---\eg, Aether~\cite{aether-deploy}, free5GC~\cite{free5gc}, Open5GS~\cite{open5gs} (\Cref{fig:cmp-trd-sota-vs-proposed}a)---by decomposing its control plane into a set of fine-grained microservices, \eg, AMF, SMF, PCF, PRF, and more (\Cref{sec:app-5gc-funcs}), connected via a service-based interface (SBI), \eg, HTTPS/REST.
The 3GPP specification~\cite{3gpptr23.501} defines each of these microservices and the functions (and features) they expose to other services (\eg, network slicing) over the shared SBI. 
The user-plane elements (\eg, UPF~\cite{5gsysapp}), however, still operate as custom-built, high-performance (DPDK- or switch-accelerated) processes to sustain multi-Tbps data traffic~\cite{macdavid2021p4, zhang2021high}.

%% file: sections/motivation.tex
\section{Motivation \& Challenges}
\label{sec:motivation}

With the evolving mobile network landscape (5G and beyond), the need to balance flexibility and scalability without compromising performance is becoming increasingly urgent (\S\ref{ssec:challenges}). 
Service-Based Architecture (SBA) cores promise modularity, but their disaggregated nature introduces new latency and scaling challenges (\S\ref{ssec:insights}).

\subsection{Balancing Flexibility \& Scalability for Performance in SBA Mobile Cores}
\label{ssec:challenges}
While the SBA-based core provides greater flexibility and scalability, its disaggregated nature and inter-communication overhead can compromise end-to-end performance (\ie, latency)---especially when coupled with the recent rise in the number of 5G subscriptions, small cell sizes (increasing handoffs), and new classes of user equipment or UEs (\eg, connected cars and massive IoT devices), each with varying performance demands~\cite{massiveIot}.
The volume of events can exceed tens of thousands of requests per second with latency service-level objectives (SLOs) of as low as \SI{20}{ms}~\cite{imt2020, etsi}.
Sprint, for example, required supporting 5,000 requests/second when deploying a 5G core~\cite{sloRequirnment}.

\begin{figure*}[tp]
    \centering
     \begin{subfigure}[b]{0.245\textwidth}
         \centering
         \includegraphics[width=\textwidth]{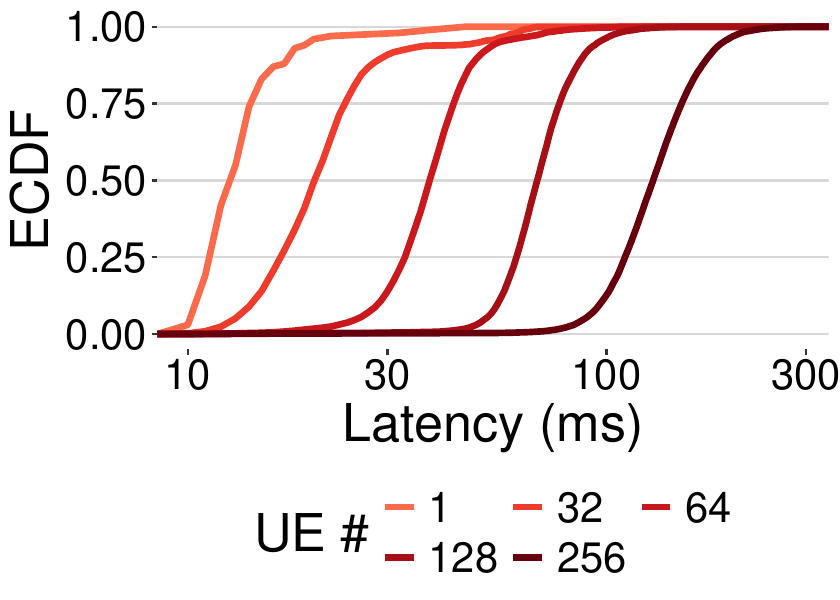}
         \vspace{-0.25in}
         \caption{REG}
         \label{fig:lat-reg}
     \end{subfigure}
    \begin{subfigure}[b]{0.245\textwidth}
         \centering
         \includegraphics[width=\textwidth]{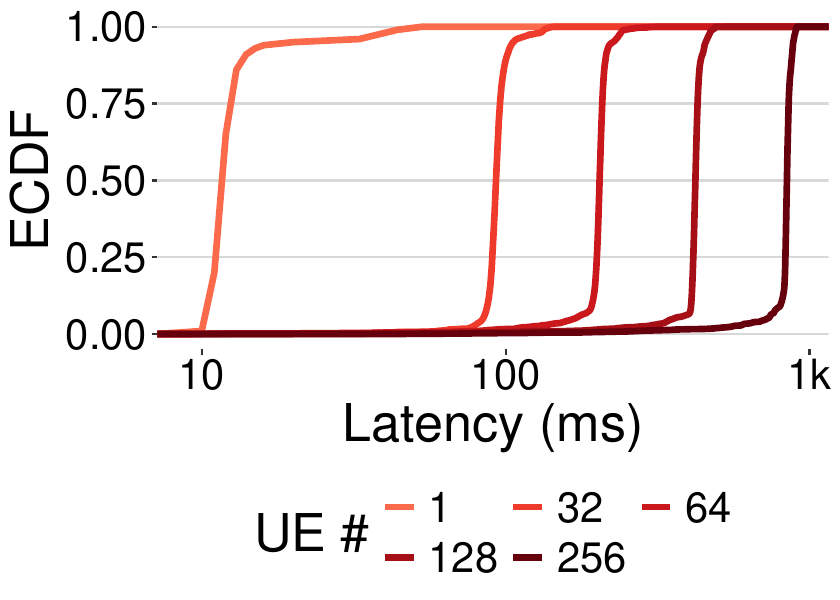}
         \vspace{-0.25in}
         \caption{PDU}
         \label{fig:lat-pdu_sess}
     \end{subfigure}
     \begin{subfigure}[b]{0.245\textwidth}
         \centering
         \includegraphics[width=\textwidth]{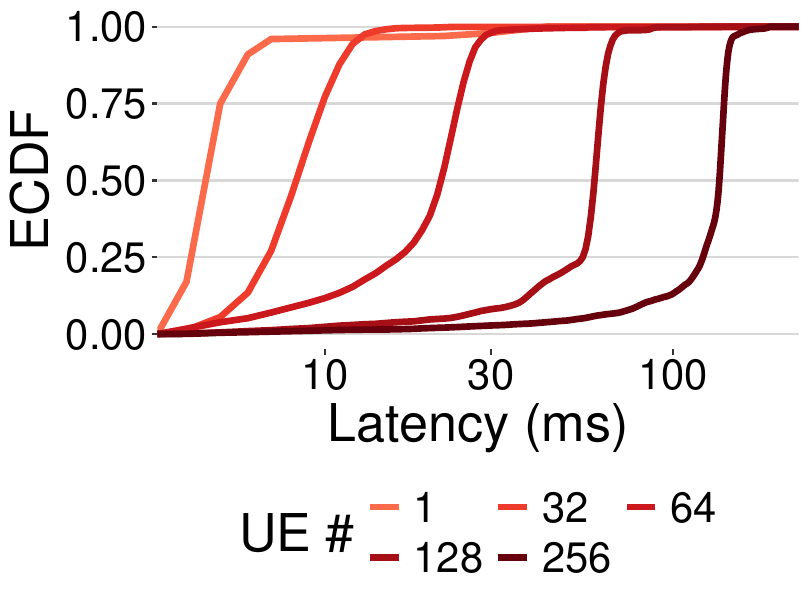}
         \vspace{-0.25in}
         \caption{SR}
         \label{fig:lat-srv_req}
     \end{subfigure}
     \begin{subfigure}[b]{0.245\textwidth}
         \centering
         \includegraphics[width=\textwidth]{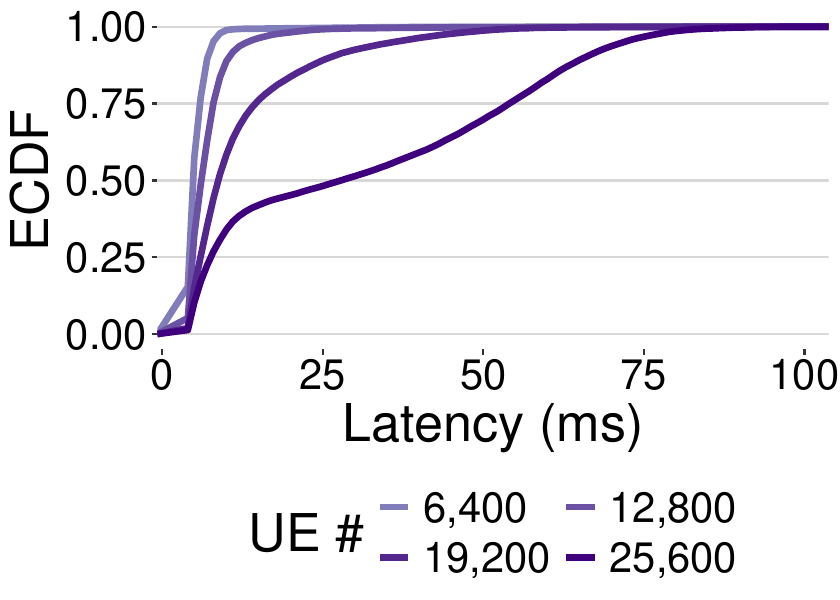}
         \vspace{-0.25in}
         \caption{SR (Real Trace)}
         \label{fig:lat-real}
     \end{subfigure}
     \vspace{-8pt}
    \caption{The end-to-end latency of different control events using synthetic (a, b, c) and real (d) traces in Aether~\cite{aether-deploy}.}
    \label{fig:cp_lat}
	\vspace{-12pt}
\end{figure*}

There seems to be a struggle between flexibility and scalability when striving for higher performance (\Cref{tab:core-comparison})---to get one, you have to sacrifice the other.
{\em But does this mean we should abandon the effort to disaggregate mobile core into microservices and instead revert to monolithic, NFV-based architectures?
We believe the answer is no.}

Current SBA-based mobile cores, such as Aether~\cite{aether-deploy}, free5GC~\cite{free5gc}, and Open5GS~\cite{open5gs}, already achieve 3GPP's \SI{20}{ms} latency target for a single user equipment (UE)~\cite{imt2020, etsi}. 
For instance, as shown in \Cref{fig:cp_lat}, Aether maintains an average latency of under \SI{20}{ms} for Registration (REG), PDU session establishment, and Service Request (SR) events when serving a single UE using both synthetic (back-to-back) and real traces.\footnote{\S\ref{sec:evaluation} describes our testbed and the real-world traces.}
However, the main issue arises when scaling to a larger number of UEs, as these cores struggle to maintain latency targets. 
Aether's latency begins to rise sharply once the number of UEs reaches 32, with the latency of PDU events soaring to \SI{1000}{ms} (\Cref{fig:cp_lat}b). 
Similarly, when tested using real traces, it exceeds the latency constraints at just around 20,000 UEs (\Cref{fig:cp_lat}d), despite being expected to support millions---on par with a typical single LTE EPC~\cite{uenum-per-epc}.

\begin{figure}[t]
\centering
\includegraphics[width=0.75\linewidth]{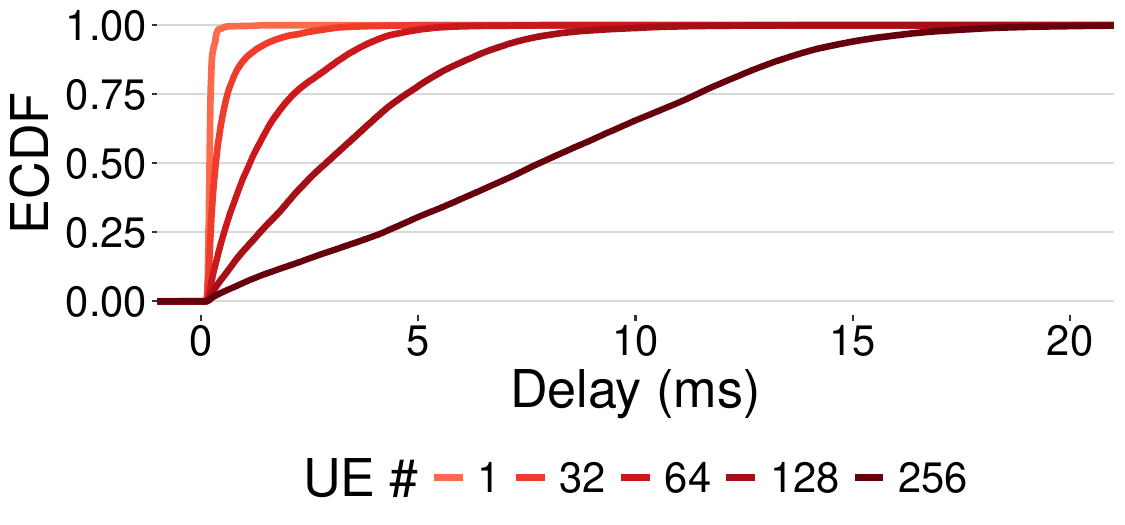}
\vspace{-10pt}
\caption{Inter-service delay in Aether: AMF $\leftrightarrow$ UDM.}
\vspace{-18pt}
\label{fig:amf-udm-request-delay-aether}
\end{figure}

One potential approach is {\em vertical scaling}, which preserves flexibility by increasing CPU and memory allocations for each service (\Cref{fig:cmp-trd-sota-vs-proposed}a). 
However, our analysis of Aether and Open5GS (implemented in Golang and C/C++, respectively) shows that resource availability is not the limiting factor. 
With 256 UEs, Aether's CPU usage peaks at just 9.96 cores (out of 80 allocated), while memory consumption remains low at \SI{0.4}{GB} (out of \SI{72}{GB}). 
Despite ample resources, UE latency still degrades, indicating that the real bottleneck lies elsewhere---\ie, congestion in inter-service communication.
For instance, measuring inter-service delay between AMF and UDM in Aether (\Cref{fig:amf-udm-request-delay-aether}) reveals a latency spike of an order of magnitude when the UE count rises from 1 to 256.

To address this, {\em horizontal scaling} offers a more viable path, increasing the number of service instances and distributing workloads across them.
Recent state-of-the-art approaches, such as L$^2$5GC~\cite{jain2022l25gc} and CoreKube~\cite{ferguson2023corekube}, take this approach but incorporate elements from past NFV-based designs---consolidating and aggregating logic to minimize inter-service communication delays.
For example, L$^2$5GC~\cite{jain2022l25gc} collocates all 5G core services on a single server or VM using a shared-memory interface, while CoreKube~\cite{ferguson2023corekube} refactors these services into a stateless monolithic application (written in C/C++) running inside a container with a separate persistent storage service (\Cref{fig:cmp-trd-sota-vs-proposed}b).
By reducing communication overhead, these solutions improve performance (and resilience), but they do so at the cost of flexibility.

Rather than sacrificing flexibility, in this paper, we argue for a fresh perspective on mobile core design---one that retains all key attributes: flexibility, scalability, performance, and resilience (\Cref{fig:cmp-trd-sota-vs-proposed}c). 
By revisiting fundamental design principles and leveraging domain insights (\S\ref{ssec:insights}), we demonstrate that to meet growing demands, it is possible to architect a 3GPP-compliant, SBA-based mobile core without compromise (\S\ref{sec:design}).

\begin{figure*}[t]
\centering
\includegraphics[width=0.9\linewidth]{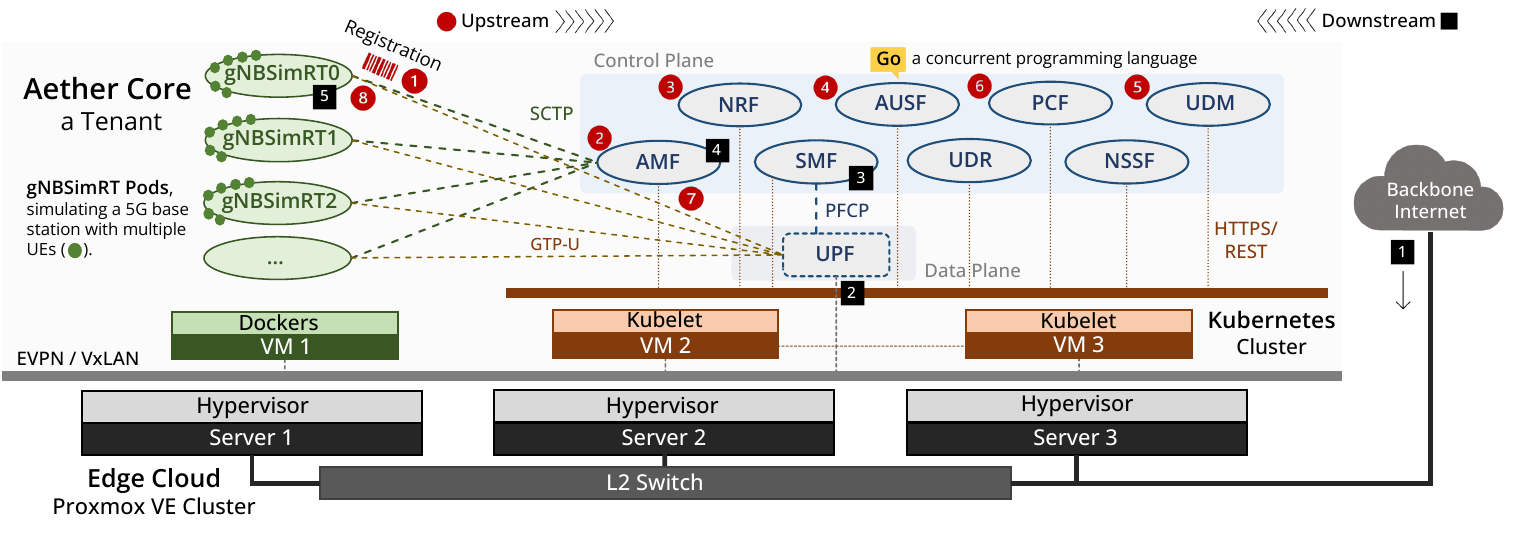}
 \vspace{-8pt}
\caption{Overview of an SBA mobile core ecosystem (Aether~\cite{aether-deploy}) deployed on our on-premise edge cloud: a Proxmox VE-based hypervisor (bottom), a cellular tenant running the Aether core as a Kubernetes workload (top-right), and a custom gNBSimRT 5G testing tool (\S\ref{sec:evaluation}) for emulating UE behavior (top-left).}
\label{fig:analysis-setup}
 \vspace{-14pt}
\end{figure*}

\subsection{Insights \& Challenges in SBA Cores}
\label{ssec:insights}
A mobile core, including 5G and 6G, occupies a distinct position within the end-to-end Internet ecosystem, sitting between UEs and the broader Internet (\Cref{fig:analysis-setup}).
It facilitates communication between end-user {\em clients} (UEs) and {\em server} applications hosted in remote data centers (such as Meta, Google, or OpenAI) or public clouds (like Microsoft Azure or Amazon AWS).
This intermediary role presents several new insights and unique design challenges ({\bf \textsf{C1--4}}) when building and deploying SBA-based mobile cores.

\vspace{2pt}
\noindent\textbf{$\bullet$ {\textsf{C1}:} Bidirectional Message Initiation and Stateful Processing in Mobile Cores.}
SBA-based mobile cores differ from traditional microservices workloads in how they process and schedule control events and messages between UEs and service instances. 
In conventional microservices architectures, message requests are predominantly unidirectional---originating from the UE and directed toward an application~\cite{aether, free5gc, open5gs}.
However, in mobile cores, communication is inherently bidirectional (\Cref{fig:cmp-trd-sota-vs-proposed}a). 
Messages can be initiated upstream, from the UE to the core over an SCTP channel via a Radio Access Network (RAN)---\eg, when a UE exits the Idle state, \Cref{fig:analysis-setup}: \ballnumber{1--8}---or downstream, from the Internet to the core and then to the UE over a PFCP channel via a UPF---\eg, an incoming voice call, \Cref{fig:analysis-setup}: \boxtext{1--5}.
To ensure consistency, these messages must be handled by the same service instance in a stateful manner---introducing challenges, for example, when using stateless processing with random load balancing~\cite{ferguson2023corekube}.
This stands in stark contrast to the conventional microservices architectures (which run server applications) that primarily consume client messages from a single ingress channel, rather than initiating messages to clients themselves~\cite{rajavaram2019automation, sayfan2019hands}.

\vspace{2pt}
\noindent\textbf{$\bullet$ {\textsf{C2}:} Complex and Diverse Service Dependencies within and across SBA Cores for Different UE Events.}
In the SBA core, various UE events (\eg, Registration, PDU Session Establishment, and Service Requests) trigger different subsets of services (\eg, AMF, SMF, and NRF). 
Moreover, within each event message, these services interact in complex ways~\cite{3gpptr23.501}.
For instance, during the Registration (REG) event, a UE sends three messages to the core. 
In the initial message, AMF interacts with NRF, followed by AUSF, which then communicates with UDM. 
In the second message, AMF connects directly with UDM, followed by PCF interacting with UDM in the third message.
Unlike conventional multi-tiered microservices architectures---where each tier communicates only with its immediate successor~\cite{aether, free5gc}---the SBA core exhibits a more complex, mesh-like communication structure.

Furthermore, SBA core components (such as AMF and SMF) serve multiple roles, including managing different network slices (\eg, for eMBB or URLLC), supporting diverse IP formats (\eg, IPv4 and IPv6), and handling events from other cores (\eg, during handovers)~\cite{3gpptr23.501, 3gpptr38.502}.
This added flexibility introduces fundamental challenges in load balancing.
Traditional service meshes, such as Linkerd~\cite{linkerd} and Istio~\cite{istio},
assume a simple model where a sender forwards requests to a service, which then distributes traffic among its instances.
In contrast, SBA core services (like AMF and SMF) have varying workloads and responsibilities, necessitating a more informed approach where the sender itself must be aware of instance load before routing requests. 
Without a more global and dynamic load-balancing mechanism, senders risk overwhelming specific instances, leading to performance bottlenecks and state inconsistencies---such as {\em Time Travel}~\cite{sun2022automatic}, where requests are processed by outdated instances that lack the latest UE state, potentially disrupting service continuity.

\begin{table}[t]
\footnotesize
\centering
\begin{tabular}{c|c}
\toprule
\textbf{Control Event} &  \textbf{Service Hotspots (\%Time)}  \\
\midrule
Registration (REG) & AMF (44.02\%), UDM (25.46\%),  \\
\hline
PDU Sess. Est. & AMF (44.55\%), SMF (30.10\%) \\
\hline
Service Request (SR) & AMF (80.02\%), SMF (19.96\%) \\
\bottomrule
\end{tabular}
\vspace{-5pt}
\caption{Top two Aether services contributing to the end-to-end UE latency for each control event (\#UE = 1).}
\label{tab:uo-e2e-ltcdb}
\vspace{-18pt}
\end{table}

\vspace{2pt}
\noindent\textbf{$\bullet$ {\textsf{C3}:} Uneven Load Distribution across SBA Core Services when Processing Different UE Event Types.}
The third challenge in SBA mobile cores lies in the uneven distribution of processing load among services when handling a specific UE event (\eg, REG, PDU, or SR). 
Certain services need to scale up rapidly compared to others to keep latency within acceptable thresholds.
However, this scaling must occur without disrupting bidirectional communication (\textbf{\textsf{C1}}).
For instance, during a Service Request (SR), the AMF experiences approximately four times the load of the SMF (\Cref{tab:uo-e2e-ltcdb}). 
This imbalance adds to the complexity of SBA core development, further complicating performance optimizations.

\vspace{2pt}
\noindent\textbf{$\bullet$ {\textsf{C4}:} Hierarchical Timeouts and Retransmission Strategies in SBA Cores and UEs.}
An SBA core must continue processing UE events despite service failures or interruptions, ensuring reliable and stable communication for both UEs and applications. 
However, maintaining this resilience while addressing \textbf{\textsf{C1--3}} poses a significant challenge.

In the event of a service instance failure, the core must seamlessly reallocate the UEs previously managed by that instance. 
One strategy for improving resilience is to leverage the hierarchical timeout and retransmission mechanisms in an SBA core, particularly in 5G and 6G~\cite{3gppts24.501}.
Unlike TCP, which relies on a flat exponential backoff strategy before terminating a connection, an SBA mobile core employs a structured escalation model: failures at lower levels---such as a Service Request (SR) timeout (\textsf{T3550})---trigger retries at progressively higher layers, including PDU Session and Registration (REG) timeouts (\textsf{T3580} and \textsf{T3510}, respectively)~\cite{3gppts24.501}.
This hierarchical approach can enhance failure recovery by enabling selective reattempts, reducing unnecessary connection terminations, and ensuring session restoration (\S\ref{sec:design}).

Additionally, stateful services (like UDSF/UDR) in SBA cores can help maintain session continuity. 
In contrast to TCP, which lacks built-in state recovery and typically resets connections upon failure, SBA cores in 5G and 6G can retain and restore session states, making them inherently more resilient to network disruptions. 
When utilized effectively, these mechanisms can improve service availability and robustness in SBA cores during failures.

%% file: sections/design.tex
\section{Design of \name{}}
\label{sec:design}

We now present the design of \name{}, a fast, flexible, scalable, and resilient SBA mobile core that addresses challenges {\bf \textsf{C1--4}} (\S\ref{sec:motivation}) while achieving the following goals: 
(1) {\em Flexibility:} Supports seamless integration of new services from different developers and vendors. For example, an operator could deploy AMF from Nokia, SMF from Cisco, and other services~\cite{mateus2021security,disaggregation-5g-services}. 
(2) {\em Scalable Performance:} Maintains high performance (\eg, low latency) as the number of UEs grows.
{\em Fault Tolerance:} Ensures resilience against failures with minimal disruption to UE connectivity.
Most existing implementations meet only a subset of these goals (\Cref{tab:core-comparison}).

\vspace{2pt}
\noindent\textbf{Design Considerations and Overview.}
\name{} enhances scalability, resiliency, and efficiency while maintaining the flexibility of SBA-based mobile cores.
Rather than relying on a fully stateless model, it employs a soft-state approach with scalable microservices patterns, where per-UE context is retained within service instances and asynchronously committed to persistent storage. 
This approach minimizes storage I/O overhead by preventing frequent writes from delaying UE message processing.
In the event of a failure, a newly spawned instance performs a synchronous read to retrieve the latest committed state, but only after pending writes have been completed. 
By shifting storage overhead from steady-state writes to recovery-time reads, \name{} reduces operational impact while enabling seamless fault recovery.

A critical challenge in SBA cores is ensuring consistent UE state across bidirectional event flows ({\bf \textsf{C1}}), where both upstream (UE-initiated) and downstream (network-initiated) messages must be handled by the same instance. 
Without a mechanism to enforce this, state inconsistencies can arise, leading to failures or crashes. 
\name{} addresses this with UE-based sticky load balancing, ensuring that all control messages for a UE are routed to the same instance. 
Doing so not only enhances system stability but also reduces storage lookups, as instances rely on the soft state rather than repeatedly querying persistent storage.

Moreover, unlike traditional microservices, SBA services must handle complex, multi-functional workloads, such as managing network slices, supporting various IP formats, and processing inter-core handovers ({\bf \textsf{C2}}).
Consequently, service-local request forwarding---used in service meshes like Linkerd~\cite{linkerd}---is inadequate, requiring services to have global, real-time visibility into instance load across the system to make informed routing decisions. 
\name{} addresses this challenge through distributed state sharing, enabling services to exchange load information to scale service instances and efficiently balance traffic, preventing bottlenecks.

By integrating scalable microservices design patterns and resilient per-UE state management (\S\ref{ssec:patterns}) with UE-based load balancing and autoscaling (\S\ref{ssec:au-lb}), \name{} delivers high performance, fault tolerance, and efficient resource utilization (\S\ref{ssec:res-ft}) while maintaining the modularity (\S\ref{ssec:flex-agile}) of SBA.

\subsection{\name{}'s Key Architecture Components}
\label{ssec:patterns}

\subsubsection{Scalable SBA Microservices Design Patterns}
\label{sssec:ss-abstractions}

\name{} introduces four different microservice design patterns for implementing the mobile core services, {\bf \textsf{P1--4}} (\Cref{fig:service-patterns}).

\begin{itemize}[leftmargin=*]
\item \vspace{-4pt}{\bf \textsf{P1: Gateway.}} 
In Aether~\cite{aether}, free5GC~\cite{free5gc}, Open5GS~\cite{open5gs}, a single microservice handles both SCTP connections from UEs (via RAN) and AMF services.
Similarly, the SMF terminates PFCP connections from the UPF (\Cref{fig:analysis-setup}).
\name{} decouples these functions into separate services and introduces a \textsf{Gateway} pattern {\bf \textsf{P1}} for each SCTP and PFCP channel (\Cref{fig:service-patterns}).
For example, the Gateway receives SCTP messages from the RAN and selects and forwards them to an appropriate AMF instance over the SBI interface.

\item \vspace{-4pt}{\bf \textsf{P2: Long-Running.}} 
AMF, SMF, and NRF follow the {\bf \textsf{P2}} \textsf{Long-Running} microservices pattern. 
They process incoming messages from other services and synchronize their state (\eg, per-UE context: SUPI, GUTI, and more)~\cite{3gpptr38.502} with a storage service over the SBI interface. 
Moreover, multiple instances of these services can run in parallel, each executing a long-running routine that also maintains a soft state of the UEs' context it is processing (\Cref{fig:service-patterns}).
Each instance load balances and forwards messages to other service instances via SBI.

\item \vspace{-4pt}{\bf \textsf{P3: Ephemeral.}}
Other services in SBA mobile core (\eg, UDM, PCF, and more) operate in a stateless manner; they launch transient (short-lived) routines to transform the incoming messages directly before sending them to the next service. 
\name{} implements them using the pattern, {\bf \textsf{P3}} (\Cref{fig:service-patterns}), which is similar to {\bf \textsf{P2}}, except it does not maintain a soft state and relies on \textsf{Ephemeral} routines for event handling.

\item \vspace{-4pt}{\bf \textsf{P4: Storage.}}
Lastly, {\bf \textsf{P4}} is a typical microservices pattern for hosting persistent storage services (\eg, for storing per-UE state and synchronizing it among instances).
\end{itemize}

\subsubsection{Resilient State Management.~}
\label{sssec:sm-strategies}
In addition to managing per-UE state, \name{} also tracks service-level state (\eg, number of instances of a service or the messages they handle) necessary for autoscaling and load balancing (\S\ref{sssec:as-service}) as well as failure handling (\S\ref{ssec:res-ft}).
Accessing these states can directly impact \name{}'s resiliency and performance when handling incoming messages.
To tackle this issue, in \name{}, we employ two separate strategies: (1) a persistent synchronization and caching scheme to retain per-UE state during failures and (2) a distributed (publish/subscribe-based) resource-sharing mechanism to share service-level state across instances.

\begin{figure}[t]
\centering
\includegraphics[width=0.92\linewidth]{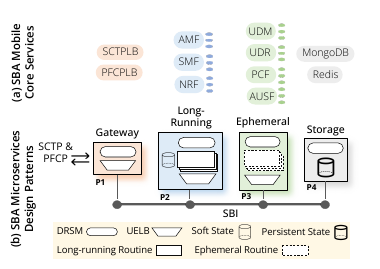}
 \vspace{-4pt}
\caption{Mapping mobile core services (a) to \name{}'s SBA-based microservices design patterns (b).}
\label{fig:service-patterns}
\vspace{-18pt}
\end{figure}

\vspace{2pt}
\noindent\textbf{Persistent per-UE State Synchronization.}
With scalable microservices pattern {\bf \textsf{P2}}, \name{} moves per-UE state (\eg, SUPI, GUTI, and more) into a storage service, as shown in~\Cref{fig:service-patterns}.
This differs from Aether~\cite{aether}, where the state is kept locally within the service itself.
While this approach increases latency when accessing the storage, we observe that the state is primarily write-only---written by the service and only read when a message for an existing UE is processed by a different instance, such as after an instance failure.
To minimize the impact on message processing, \name{} maintains a soft copy of the per-UE context within each service instance, asynchronously updating the storage. 
In the event of a failure, the new service instance retrieves the last known state from the storage, reducing the risk of retransmission or loss of the incoming control message.

\vspace{2pt}
\noindent\textbf{Distributed (Service-Level) State Sharing.}
In \name{}, multiple services (and instances) need to share continuously evolving state (\eg, message load, instance count, and unique NGAP IDs among AMFs) with each other for autoscaling, load balancing, and failure handling.
For example, when balancing incoming messages to other services' instances, the sending service needs to know the current load on these pods to distribute messages evenly.
Similarly, AMF and SMF instances need to assign unique NGAP IDs and IP addresses to incoming UEs, requiring coordination to track the allocation of these IDs and addresses.
One approach is to manage this state on a central schedular that each service instance communicates with. 
However, this communication can become a bottleneck, incurring excessive overheads when processing incoming messages.

Instead, \name{} implements a distributed (publish/subscribe) resource-sharing mechanism (DRSM) to allow different services (and instances) to synchronize their state information without having to query a central schedular.
The DRSM scheme ensures that all services maintain a globally consistent and up-to-date view of the subscribed state.

\subsection{Autoscaling \& Load Balancing}
\label{ssec:au-lb}

\subsubsection{Autoscaling Service Instances.}
\label{sssec:as-service}
The two patterns {\bf \textsf{P2--3}} allow \name{} to horizontally scale its mobile core services (\eg, AMF, SMF, and UDM) by creating new instances.
\name{} utilizes the underlying microservice runtime (\eg, Kubernetes HPA~\cite{k8s}) to track the system's CPU and memory usage of each service; it then starts new instances as the usage goes beyond a specified threshold.
{\bf \textsf{P2}} utilizes the DRSM service to retrieve a unique set of NGAP ID and IP blocks for AMF and SMF, respectively, before servicing requests.
{\bf \textsf{P3}} microservice instances (\eg, for PCF or UDM), on the other hand, do not maintain a local per-UE soft state. Instead, upon boot, they begin publishing their resource usage (\eg, UE request load) using DRSM.
Upon removal, DRSM notifies other instances about the departure of a service instance and, in the case of {\bf \textsf{P2}}, also redistributes state (\eg, the NGAP ID and IP blocks) to the remaining active instances.

Lastly, regardless of their type, each instance (de-)registers with the NRF service, per the 3GPP specification~\cite{3gpptr38.502}, which tracks information about the various services available in the core (and the functions they support) and stores it persistently in the storage service {\bf \textsf{P4}}.

\subsubsection{Load Balancing UE Messages.}
\label{sssec:lb-requests}

Existing microservices implementations for scheduling messages are insufficient for bidirectional (upstream/downstream) communication (\S\ref{sec:motivation}), as they are optimized for the unidirectional flow and route messages based on connection-level (L4, \eg, ClusterIP~\cite{k8sServices} or NodePort~\cite{k8sServices}) or RPC-level (L7, \eg, Linkerd~\cite{linkerd} or Envoy~\cite{envoy}).

In the bidirectional flow, messages for a specific UE must be handled consistently by the same service instance. 
Otherwise, this can lead to collisions and inconsistencies due to the asynchronous arrival of messages from either direction when processed across different instances.
For example, AMF can simultaneously receive messages from the RAN (over SCTP) or an SMF instance (via UPF over PFCP) for the same UE.
It can happen when AMF brings a UE out of the Idle state due to (a) some activity on the phone indicated by SR or (b) any downstream data received (\eg, incoming voice call) by the UPF indicated to the SMF (over PFCP).\footnote{AMF initiates a {\em paging} process~\cite{downlinkSmfNotification} to search the RAN with the respective UE and sends a request to open a GTP-U tunnel over the network.}

In \name{}, we address this issue by developing a custom, client-side load-balancing module, {\em UELB} (\Cref{fig:service-patterns}).
In conjunction with Kubernetes headless service~\cite{k8sServices} and the DRSM scheme, UELB efficiently schedules UE messages among the SBA core service instances while ensuring the bidirectional constraints.

\vspace{2pt}
\noindent\textbf{UELB Workflow.}
Except for storage {\bf \textsf{P4}}, the remaining service patterns {\bf \textsf{P1--3}} incorporate the UELB module (\Cref{fig:service-patterns}b).
The messages first arrive at the SCTPLB service {\bf \textsf{P1}}, which acts as a proxy between the RAN and the AMF.
At boot-up, it subscribes via DRSM for the service-level state (\eg, message load and instance count) for the AMF instances identified by the NRF service, which tracks all active service instances within (and across participating) mobile cores. 
When a message arrives, it looks up the UE identifier (NGAP ID) from the SCTP packet. 
If present, it selects the corresponding AMF instance from the local DRSM agent.
Otherwise, in case of a new request, it forwards it to an AMF instance with the least load.
From there onwards, all requests belonging to that UE arrive at the selected AMF. 
The AMF also commits to DRSM, informing other services about its ownership of the UE---doing so ensures that messages in both directions are processed by the same instance (\eg, SMF communicating with AMF for the downlink messages).

The same process follows for {\bf \textsf{P2--3}}, except the UE identifier is now looked up from the internal SBI (HTTPS/REST) message.
Moreover, as new instances arrive or leave (\eg, due to autoscaling), DRSM automatically synchronizes the state in the local agents of all subscriber services (and instances).

\subsection{Resiliency \& Fault Tolerance}
\label{ssec:res-ft}

The existing Aether~\cite{aether} and free5GC~\cite{free5gc} cores are limited to running only one (stateful) instance of a particular service. 
When that instance fails, the whole core crashes, causing Kubernetes to spawn a new instance of the failed services. 
Moreover, the UEs also fail, triggering the control event sequence to restart all the way back from Registration.

In contrast, \name{} supports multiple instances of a given service; hence, if one fails, the other can take over---allowing a UE to recover quickly without triggering failures at lower levels, \eg, avoiding an SR timeout that could have escalated to PDU session and Registration failures (\S\ref{ssec:insights}).
To enable seamless migration of UE messages between service instances (\eg, upon failures), \name{} adds support for the following: 
{\em (a) NRF keepalive}, which allows the NRF service to keep track of the liveness status of all core service instances (\eg, AMF, SMF, and UDM). 
If a service instance fails or does not respond in time (\eg, overloaded or down), NRF will take it out of circulation, waiting for the keepalive messages to start again. 
Doing so ensures that services receive an active set of target instances to select from when load-balancing the arriving messages.
{\em (b) DRSM time-to-live (TTL)}, which removes stale entries corresponding to the failed instance(s) from the local DRSM agent of the subscriber services.
It also consistently redistributes state (\eg, the NGAP ID and IP blocks) to the remaining active instances.

After the message is re-sent (upon timeout), the newly chosen instance synchronously retrieves the most recent state from storage, updates the local per-UE soft state, and continues processing the message from where it was interrupted.
In short, NRF keepalive and DRSM TTL allow services in \name{} to fail and restart without disrupting the core (\S\ref{sec:evaluation}).

\subsection{Flexibility \& Agility}
\label{ssec:flex-agile}
Lastly, \name{} retains the flexible and agile characteristics of a service-based architecture (SBA), similar to Aether and free5GC. 
It adheres to and supports modern Cloud and DevOps lifecycle management practices, especially in addressing the ongoing challenge of continuously updating services through CI/CD~\cite{rajavaram2019automation}.
Moreover, it enables modular plug-and-play integration of new mobile core service functions and ensures vendor independence by allowing the composition of services from varying vendors and developers~\cite{mateus2021security}.

%% file: sections/implementation.tex
\section{Implementation}

\name{} is designed as a scalable and fault-tolerant 5G core, implemented in Go ({\sf v1.18.3}), enabling seamless integration with Aether \cite{aether}. We introduced key enhancements by implementing Gateway, UELB, DRSM, and soft-state modules, modifying the following Aether base modules: AMF ({\sf \href{https://github.com/omec-project/amf/tree/2c46569e}{commit:2c46569e}}), SMF ({\sf \href{https://github.com/omec-project/smf/tree/20fd5cb3}{commit:20fd5cb3}}), UDM ({\sf \href{https://github.com/omec-project/udm/tree/69566599}{commit:69566599}}), UDR ({\sf \href{https://github.com/omec-project/udr/tree/35eb7b76}{commit:35eb7b76}}), AUSF ({\sf \href{https://github.com/omec-project/ausf/tree/c84dff44}{commit:c84dff44}}), PCF ({\sf \href{https://github.com/omec-project/pcf/tree/bcbdeb0c}{commit:bcbdeb0c}}), and NRF ({\sf \href{https://github.com/omec-project/nrf/tree/b747b985}{commit:b747b985}}). 
We modified \textasciitilde{}4,100 lines of code (LoCs) across these modules, with the bulk of changes in AMF (51\%), SMF (15\%), AUSF (12\%), and UDM (12\%), respectively. 

Additionally, we implemented the gateway as a new Go microservice in \textasciitilde{}2,800 LoCs, to handle incoming SCTP and PFCP connections. 
It tracks active 5G microservice instances and monitors AMF/SMF loads via the DRSM module, selecting the least-loaded AMF (by UE count) for each new UE.
For persistent UE caching, our DB backend utilizes RedisDB ({\sf v7.4.2})~\cite{redis}.
Since NRF lacks native support for publish/subscribe capability, we utilize MongoDB ({\sf v10.31.5})~\cite{mongodbchangestream} as a separate service, which provides publish/subscribe support and state persistence.

Each service instance incorporates a DRSM module (\Cref{fig:service-patterns}), which talks to the NRF module acting as a message broker (using MongoDB service)~\cite{mongodbchangestream}; other services subscribe for the state information they need.
The DRSM module relies on three key features in MongoDB: (1) {\em time-to-live (TTL)}~\cite{mongodbttl}, to discard expired state (\eg, once an instance is removed); (2) {\em atomic transactions}~\cite{mongodbcatomicoperations}, to ensure consistent access to state (\eg, AMF requesting unique NGAP IDs to allocate to UEs\footnote{In \name{}, AMF and SMF instances are initially provided with distinct blocks of NGAP IDs and IPs, respectively. Any further allocations to UEs are managed locally, requesting additional blocks as needed.}); and (3) {\em change streams}~\cite{mongodbchangestream}, to detect and notify instances about the changes in state.

%% file: sections/evaluation.tex
\section{\name{} Evaluation}
\label{sec:evaluation}

\begin{figure*}[t]
    \centering
     \begin{subfigure}[b]{0.9\textwidth}
         \centering
         \includegraphics[width=\textwidth]{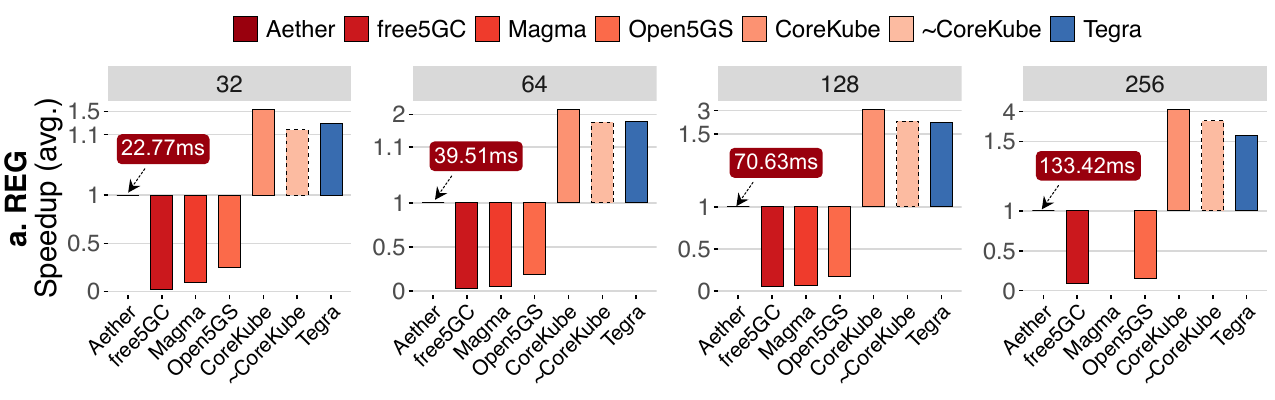}
         \vspace{-15pt}
         \label{fig:e2e-reg-core}
     \end{subfigure}
    \begin{subfigure}[b]{0.9\textwidth}
         \centering
         \includegraphics[width=\textwidth]{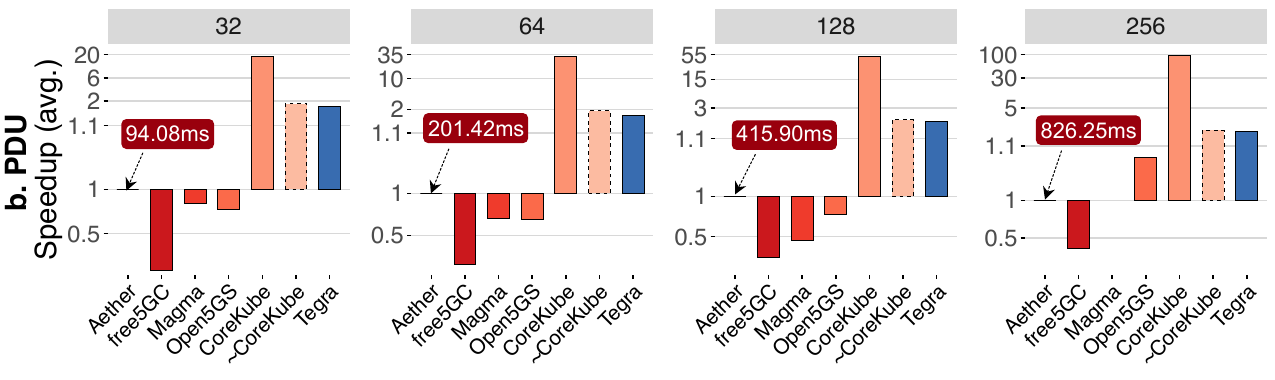}
         \vspace{-15pt}
         \label{fig:e2e-pdu-core}
     \end{subfigure}
     \begin{subfigure}[b]{0.9\textwidth}
         \centering
         \includegraphics[width=\textwidth]{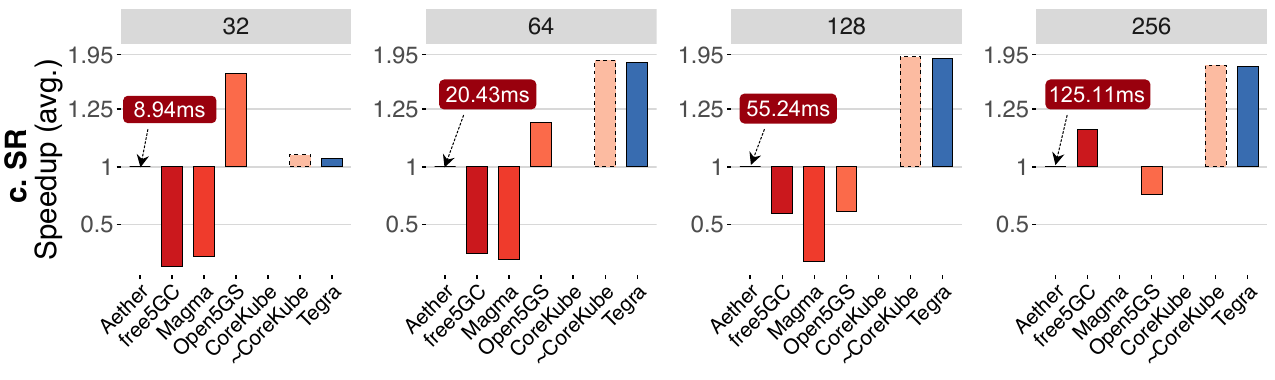}
         \vspace{-12pt}
         \label{fig:e2e-sr-core}
     \end{subfigure}
     \vspace{-12pt}
    \caption{Improvement in end-to-end control-plane latency with \name{} over existing open-source 5G cores.}
    \vspace{-15pt}
    \label{fig:e2e-speedup-with-other-cores}
\end{figure*}

\subsection{Experiment Setup}
\label{ssec:exp-setup}
\subsubsection{A Realistic Test Environment.~}
To evaluate the effectiveness of \name{}, we run experiments using the on-premises edge cloud as our test environment, shown in \Cref{fig:analysis-setup}.
It consists of a multi-tenant hypervisor and a (cellular) tenant (\eg, Aether), operating on a Kubernetes cluster~\cite{k8s}.
We also built our custom UE/RAN emulator, gNBSimRT, to emulate various control event traces with multiple UEs. 
Together, these allow us to evaluate all aspects of an SBA-based mobile core, as deployed in the real world. 

\vspace{2pt}
\noindent\textbf{A Proxmox VE-based Edge Cloud.} 
We assemble our multi-tenant edge cloud (the bottom half in \Cref{fig:analysis-setup}) using Proxmox VE ({\sf v7.4})---an open-source platform that provides complete compute, storage, and network virtualization (via KVM hypervisors)~\cite{proxmox}.
The cloud consists of three servers with Intel Gold~5220R CPU @ \SI{2.20}{GHz}~(96 cores and two sockets), connected using an Intel X710 40G NIC and a Tofino1 switch~\cite{tofino}.
Tenants run their virtual machines (VMs) over a virtual private network (VPN) based on the EVPN-VxLAN overlay technology~\cite{overlay}---common in most cloud environments~\cite{overlay1}.

With the addition of new components (\eg, gateway service), we allocate ten CPU cores to the gateway service and ten cores to the storage, which serves as a backend-as-a-service (BaaS).

Another ten cores are dedicated to managing the Kubernetes cluster, while 50 cores are allocated to handle the SBA-based mobile core workload (\name{}, Aether~\cite{aether}, Magma~\cite{magma}, CoreKube~\cite{corekube-github}, or free5GC~\cite{free5gc}). 
The remaining CPU cores are assigned to the Proxmox hypervisor.

\vspace{2pt}
\noindent\textbf{gNBSimRT: Our Custom 5G Testing Tool.}
We build an extension, called gNBSimRT, based on the open-source gNBSim testing tool~\cite{gnbsim}, which is provided as part of the Aether project---also a Docker container---to exercise the 5G control plane by emulating various UE events, \eg, Registration (REG), PDU Session Establishment, and Service Request (SR).
gNBSimRT allows users to specify different event profiles (as state machines)~\cite{gnbsim} to emulate the behavior of a UE (\ie, transitioning from REG to PDU to SR), with multiple UEs processing simultaneously, in parallel.
Moreover, it can generate real traces using a user-specified traffic distribution~\cite{meng2023modeling}.

\subsubsection{Mobile Core Baseline Systems.~}
Our evaluation compares \name{} against five representative 5G mobile core frameworks: Aether~\cite{aether}, Magma~\cite{magma}, Open5GS~\cite{open5gs}, free5GC~\cite{free5gc}, and CoreKube~\cite{corekube-github}. 

Aether, Magma, free5GC (implemented in Golang~\cite{golang}), and Open5GS (written in C) follow a microservices-based architecture, which provides modularity and scalability suited for cloud-native deployments. 
On the other hand, CoreKube is based on Open5GS libraries (written in C) and consolidates core functions into a single monolithic binary.
CoreKube’s monolithic approach aims to minimize inter-service communication overhead, offering potential performance benefits at the cost of flexibility.

However, as of now, CoreKube~\cite{corekube-github} is an incomplete 5G core implementation, lacking {\em design} support for several critical control-plane procedures required for real-world deployments, such as SR as well as UPF support in PDU~\cite{corekube-github-missing-functionality}.
To realistically estimate CoreKube's performance in practical scenarios, we construct an augmented baseline, which we refer to as \textasciitilde CoreKube. 
We estimate the additional lines of code (LoCs) required to implement the missing functionality based on the Open5GS codebase. 
To support the three control plane events used in our evaluations, REG, PDU, and SR, we estimate 67\%, 71\%, and 733\% additional LoCs, respectively, in \textasciitilde CoreKube.
With these additions, we project the associated runtime overhead by scaling CoreKube's measured latency for each event using a \SI{2.2}{GHz} CPU. 

For completeness, we present \textasciitilde CoreKube’s projected results alongside the other baselines. 
We exclude L$^2$5GC~\cite{jain2022l25gc} from our evaluation because its public artifact supports only a single UE and does not handle multiple UEs, making it infeasible for our analysis.

\subsubsection{Evaluation Metrics and Traces.}
We evaluate UE-level and system-level performance of each mobile core.
For the UE evaluations, we measure the end-to-end latency (\ie, the time an event takes when traversing mobile core services and back) as the number of UEs grow.
Moreover, we use OpenTelemetry (OTel)~\cite{otel} to further track the time taken by each service (AMF or SMF) when processing a given event along the path. 

For system-level performance, we use Prometheus~\cite{prometheus} to collect resource usage metrics (\ie, CPU and memory) and the SBI traffic between services on the Kubernetes cluster.

For event generation, we use two different traces: synthetic and real. 
In the synthetic case, each UE generates control events in a closed-loop manner (sending a request and waiting for a response before immediately sending the next one). 
In the real case, UE now waits for a certain time before sending the next event; we sample this time (uniformly at random) from a distribution of inter-arrival times (IATs), collected separately for each event from a week-long trace (provided by a large mobile operator). The trace covers about 40,000 randomly-sampled UEs~\cite{meng2023modeling}.
All the UEs operate in parallel as separate Goroutines within a gNBSimRT docker.

\subsection{End-to-End Performance}
\label{ssec:e2e-latency-res}

\Cref{fig:e2e-speedup-with-other-cores} shows the speedup for different 5G core implementations
across varying numbers of UEs (32, 64, 128, and 256) for Registration (REG), PDU Session Establishment (PDU), and Service Request (SR) events. 
The speedup is calculated relative to Aether taken as the baseline (indicated $=1$).

Results show that Magma~\cite{magma}, Open5GS~\cite{open5gs}, and free5GC~\cite{free5gc}, demonstrate higher latency and thus poorer performance compared to the baseline Aether across all three events with greater than 128 UEs. 
Open5GS outperforms Aether with SR event for less than 128 UEs but looses its advantage with higher UE counts.

Among all evaluated 5G cores, \name{} consistently outperforms Aether, free5GC, Magma, and Open5GS, exhibiting lower end-to-end latency across all three events with more than 32 UEs. 
Specifically, \name{} achieves up to 1.74$\times$ speedup (74\% improvement) for Registration, 1.5$\times$ (50\% improvement) for PDU session establishment, and 1.75$\times$ (75\% improvement) for Service Request at 256 UEs compared to Aether.
CoreKube, however, achieves even lower latency primarily due to its incomplete implementation of control-plane event functionality. 
\textasciitilde CoreKube closely matches the performance of \name{} but at the expense of flexibility.

\begin{figure}[t]
\centering
	\includegraphics[width=\linewidth]{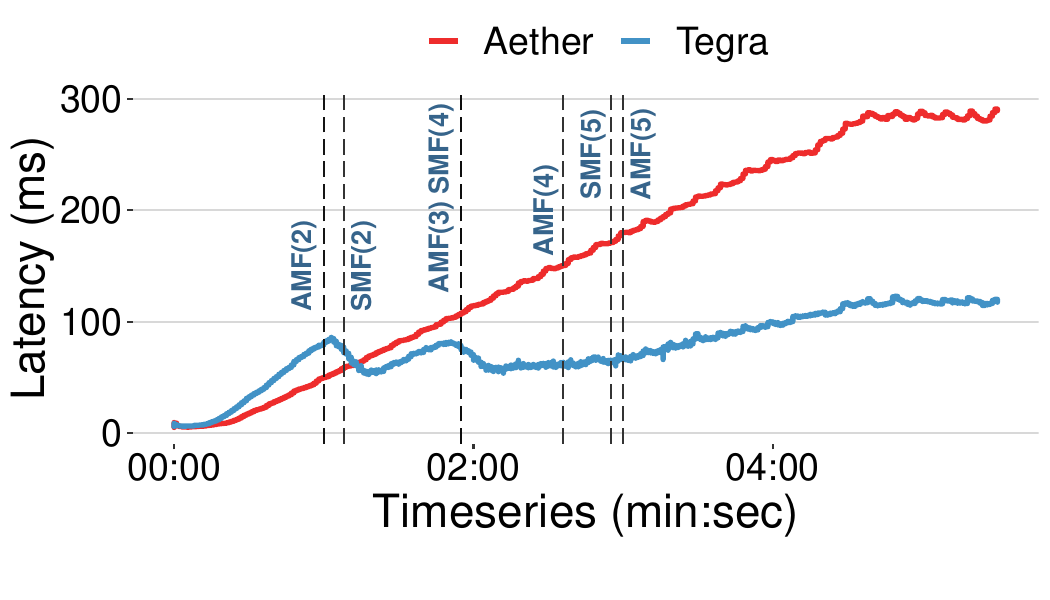}
	\vspace{-34pt}
	\caption{SR time-series in \name{} with fine-grained autoscaling; vertical lines indicate the addition of new NF instances over time.}
	\label{fig:sr_hpa_timeseries}
    \vspace{-6pt}
\end{figure}

\subsection{Microbenchmarks}
\label{ssec:micro}

\subsubsection{\name{} Efficiently Manages Load and Reduces Latency.~}
\label{ssec:autoscaling}
\Cref{fig:sr_hpa_timeseries} compares \name{}'s autoscaling feature with Aether with a time series of events and their associated latencies.\footnote{Static scaling with increasing UEs shows similar trends (\Cref{sec:static-scaling}).}
We utilize k8s Horizontal Pod Autoscaling (HPA) support with core pinning (following best practices).
The configuration includes a maximum of 4 cores per service instance, with the desired CPU utilization set to 60\% of the requested cores ($\sim$2.5 cores).
Additionally, we cap the number of instances per service at 5.
We then start sending traffic from the UEs, gradually increasing the load by introducing a new UE every second until it reaches 256.
Each UE continuously sends SR messages in a closed-loop manner, maintaining a \SI{10}{ms} gap between subsequent requests.

We observe that Aether, despite being vertically scaled (\ie, able to utilize all CPU cores), still experiences increasing latency for SR as more UEs enter the system.
On the other hand, \name{} maintains 2$\times$ lower latencies with the same number of CPU resources by spinning up new instances for AMF and SMF, which are the dominant services for SR. 
The initial high latency in \name{} is caused by the addition of SCTPLB and the database, but once more NF instances are up and running, the overall latency reduces.

\begin{figure}[t]
    \centering
     \begin{subfigure}[b]{0.23\textwidth}
         \centering
         \includegraphics[width=\textwidth]{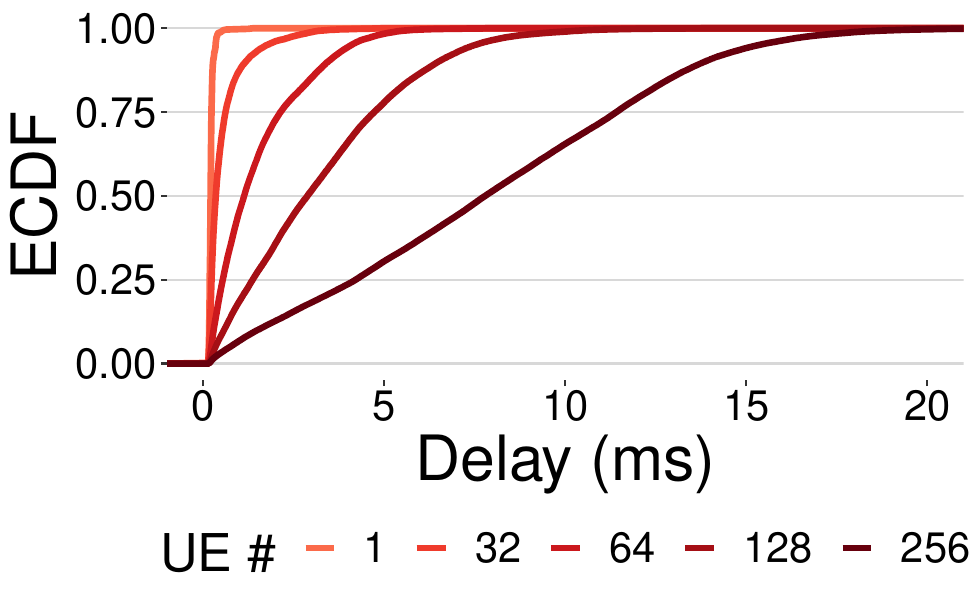}
         \vspace{-18pt}
         \caption{Aether}
         \label{fig:aetherplus-srvreq-go-latency-amf}
     \end{subfigure}
    \begin{subfigure}[b]{0.23\textwidth}
         \centering
         \includegraphics[width=\textwidth]{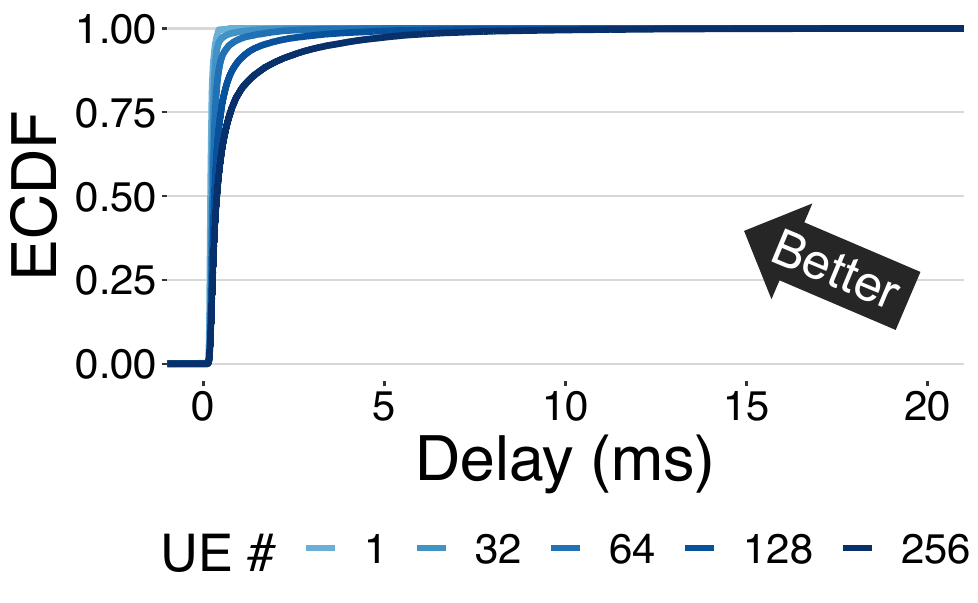}
         \vspace{-18pt}
         \caption{\name{}}
         \label{fig:aetherplus-srvreq-go-latency-smf}
     \end{subfigure}
     \vspace{-8pt}
    \caption{Inter-service delay for REG: AMF $\leftrightarrow$ UDM.}
     \vspace{-15pt}
    \label{fig:amf-udm-request-delay-tegra}
\end{figure}

\begin{figure}[t]
\centering
	\includegraphics[width=0.95\linewidth]{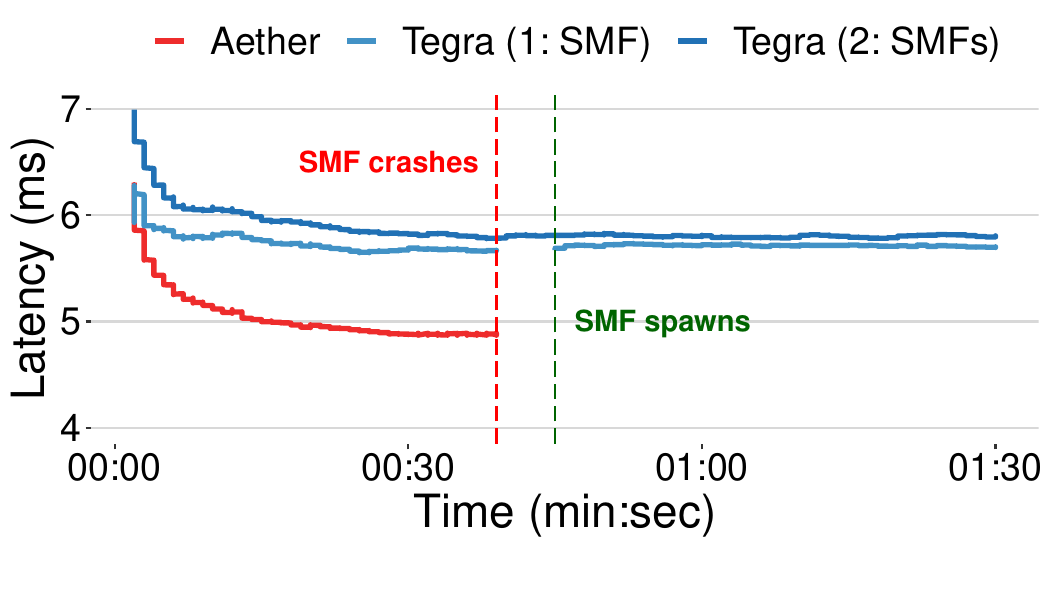}
	\vspace{-20pt}
	\caption{Comparison of failure handling between Aether and \name{} for SR.}
    \label{fig:sr_fault_tolrent2}
    \vspace{-14pt}
\end{figure}

\subsubsection{\name{} Reduces Inter-Service Delay via Finge-Grained Autoscaling.~}
\Cref{fig:amf-udm-request-delay-tegra} illustrates the impact of increasing service instances on transmission delay between network functions. 
Deploying additional instances (4 in this case) reduces the transmission delay up to a factor of 10 for 256 UEs, highlighting Tegra's effectiveness in alleviating congestion at ingress points.

\subsubsection{\name{}'s Per-UE State Caching and DRSM Effectively Minimize Latency During High Loads.~}
The per-UE state cache in \name{} reduces database calls from over 10 to just 1.1 (avg.) per event, significantly minimizing delays and cutting access times by \SI{22}{ms}, \SI{37}{ms}, and \SI{57}{ms} for REG, PDU, and SR events, respectively, for 256 UEs. 
Additionally, DRSM prevents latency spikes during high churn or failures. 
Without DRSM, overhead increases by an average of \SI{5.6}{ms} $\pm$ \SI{2.4}{ms} per UE.

\subsubsection{\name{} is Resilient to Failures, Ensuring Consistent Performance.~}
Another advantage of \name{}, over existing cores is its ability to keep operating under failure without crashing the core. 
For example, \Cref{fig:sr_fault_tolrent2} compares Aether against \name{} (with single and multiple instances). 
At the 40-second mark, we deliberately delete the SMF service using the \verb|kubectl| command, leading to Aether losing all UE state and causing UEs to time out. 

For \name{}, we conduct two tests. 
First, we operate a single instance of SMF and delete it around the same 40-second mark. 
The new SMF instance retrieves the previous state upon boot and begins processing the UE messages. 
During this period, AMF continues retrying (at 3-second intervals) until the incoming SMF processes it around the 45-second mark, which occurs within 6 seconds. 
Next, we deploy multiple instances of SMF in \name{} and delete one again at the 45-second mark. 
In this case, another instance quickly picks up the affected UE message, retrieves its state from the DB, and begins processing it immediately.

\subsubsection{\name{} Simplifies Deployment with Minimal Code Changes, Offering Maximum Flexibility.~}
We compare the flexibility of \name{} against CoreKube~\cite{ferguson2023corekube} and measure the deployment and development complexity of adding a new feature (service or event) in both.
We used two metrics to evaluate the cost: lines of code (LoC) and files changed.

Adding a new service (\ie, NRF) entails changing a single line in \name{}'s configuration file, specifying the new container image to use.
In contrast, the same task in CoreKube requires 1,705 LoCs spread across 18 files.
It is because, unlike \name{}, CoreKube implements all servers as a single monolithic application (in C/C++) inside a container, necessitating changes across the entire codebase.

Likewise, incorporating support for a new event (\ie, PDU session management) in \name{} involves simply utilizing a different image with PDU support (a single-line change in the configuration file); it deploys it using CI/CD~\cite{rajavaram2019automation}, automatically.
Meanwhile, CoreKube demands 712 modifications in six files and a complete recompilation of the core to generate a new image before deployment.
L$^2$5GC and Magma incur similar costs, which are not shown here.

\begin{table}[tp]
\centering
\footnotesize
\begin{tabular}{c|c|c|c}
\toprule
\multirow{2}{*}{\textbf{Event}} & \multicolumn{2}{c|}{\textbf{Latency (ms) / \#DB Reqs (avg.)}} & \multirow{2}{*}{\textbf{Speedup}} \\
  &  {\em With UE Cache} & {\em Without UE Cache}  \\ 
\midrule               
REG & 76.27 / 1.01 & 102.21 / 10.00 & 34.01\% \\ 
PDU & 548.52 / 1.01 & 585.12 / 12.00 & 6.67\% \\ 
SR  & 71.31 / 1.01 & 128.51 / 12.00 & 80.21\% \\ 
\bottomrule
\end{tabular}
\vspace{-5pt}
\caption{Latency and database request analysis with and without UE cache for 256 UEs.}
\label{tab:component-analysis}
\vspace{-17pt}
\end{table}

%% file: sections/related-work.tex
\section{Related Work}
\label{sec:related}

\vspace{2pt}
\noindent\textbf{Mobile Network Performance Characterization.}
Characterizing mobile network traffic has been a topic of interest for decades, and most characterization works focused on the user plane~\cite{xu2011cellular, pereira2017experimental, aceto2021characterization, jana2013network}.
Xu~\ea~\cite{xu2011cellular} characterized the cellular data network infrastructure of four major cellular carriers within the US, and showed how their findings can help with mobile-content placement and content-server selection.
Aceto~\ea~\cite{aceto2021characterization} analyzed mobile-app traffic available in the public dataset (MIRAGE-2019), adopting two related modeling approaches based on Markov models.   
However, recently, the community has started looking into the SBA mobile core's control plane.  
Xu~\ea~\cite{xu2020understanding} measured user-observed performance, including physical layer signal quality, end-to-end throughput, latency, quality of experience of 5G's niche applications, and energy consumption.  
Yuan~\ea~\cite{yuan2022understanding} studied end-to-end performance, radio access network, and energy consumption from the content provider's perspective.  
Narayanan~\ea~\cite{narayanan2021variegated} studied the performance, power consumption, and application quality-of-experience (QoE) of commercial 5G networks in the US.
These works are orthogonal to our effort as we focus on characterizing the mobile control plane.

\vspace{2pt}
\noindent\textbf{Mobile Core Optimizations.}
Due to the growing demand for low-latency services and improved scalability in recent years, the mobile core's control plane has gained increasing interest.
To optimize EPC, Li~\ea~\cite{li2017control} proposed DPCM to reduce data access latency caused by sequential control event procedures through parallel processing and managing device-side state replica.
Shah~\ea~\cite{shah2020turboepc} developed TurboEPC to offload part of the user states and control-plane procedures to programmable switches.
Qazi~\ea~\cite{qazi2017high} proposed PEPC, consolidating the UE state and re-designing EPC functions for efficient access.
Ahmad~\ea~\cite{ahmad2020low} designed Neutrino to reduce the control-plane latency by using customized serialization.  
L$^2$5GC~\cite{jain2022l25gc} and CoreKube~\cite{ferguson2023corekube} re-architects the 5G core network for event processing by consolidating network functions into a single VM or container, aiming to reduce control plane latency.
However, unlike \name{}, these approaches apply past NFV-based techniques (\eg, consolidating mobile core services), hurting flexibility.

\vspace{2pt}
\noindent\textbf{Network Function Scaling.}
Scaling network functions has been discussed in the context of EPC, and most prior works have focused on scaling MME in 4G~\cite{qazi2017high, shah2020turboepc}.
Amogh~\ea~\cite{amogh2017cloud} proposed an architecture that dynamically scales up and down required microservices for load balancing.  
Banerjee~\ea~\cite{banerjee2015scaling} proposed SCALE to scale a software LTE~MME to handle rising signaling traffic.  
It handles the load from a growing number of devices by multiplexing them across data centers.  
An~\ea~\cite{an2012dmme} proposed DMME to split the control-event procedures among multiple servers and store the UE state in an external database.
Nagendra~\ea~\cite{nagendra2019mmlite} proposed MMLite, a load-balancing solution for MMEs, leveraging skewed consistent hashing to distribute incoming connections more efficiently.
Compared to these works, \name{} provides a flexible and scalable implementation of a 3GPP-compliant SBA-based mobile core. 
With automatic load balancing and careful management of states (via per-UE caching and DRSM), \name{} ensures consistent access to the state and meets latency requirements for a large population of UEs. 

%% file: sections/conclusion.tex
\section{Conclusion}
In this paper, we highlight the conflict between two SBA-based mobile core optimization strategies: {\em disaggregation} for flexibility and {\em consolidation} for performance.
We show that consolidation is not an effective approach for optimizing mobile cores, as it contradicts service-based architecture (SBA) principles and deviates from 3GPP specifications.
Using a real on-premises edge cloud, we provide a detailed comparison of existing mobile core frameworks (\eg, Aether, free5GC, Open5GS, Magma, and CoreKube) with \name{}.
Our experiments debunk common myths, uncover new insights, and reveal that current implementations fail to meet the strict latency requirements of modern mobile use cases.
To address these challenges, we propose and implement \name{}, a fast, flexible, scalable, and resilient 3GPP-compliant SBA core designed to efficiently scale mobile control-plane services.

\vspace{2pt}
\noindent\textbf{\em Looking Ahead:}
While \name{} marks a significant step forward, our results also point to broader opportunities for improving SBA-based mobile core scalability and performance. One key area is {\em hardware and software co-optimization.} 
Increasing the number of core service instances helps mitigate synchronization block time, but execution time still dominates latency. Unfortunately, software-level improvements alone fall short in addressing this.

Recent advances show that offloading specific functions to programmable hardware can yield substantial performance gains. 
The microservice-based design of the mobile core presents a unique opportunity to intelligently combine CPUs and ASIC-based SmartNICs---equipped with hundreds of lightweight RISC-like cores---to accelerate control-plane services and meet stringent latency targets~\cite{choi2020lambda}.

We see the development of domain-specific architectures that expose a simple, event-driven programming model as a promising direction—empowering mobile operators to efficiently compose and deploy services on SmartNICs and other programmable platforms.

%% file: sections/appendix.tex
\appendix

\input{sections/app-5g-core-funcs}

%% file: sections/app-5g-core-funcs.tex
\section{Key 5GC Network Functions}
\label{sec:app-5gc-funcs}

\begin{table*}[t]
\footnotesize
\centering
\begin{tabular}{p{120pt}|p{190pt}|p{160pt}}
\toprule
\multicolumn{1}{c|}{\textbf{Component}} & \multicolumn{1}{c|}{\textbf{Description}}                                                   & \multicolumn{1}{c}{\textbf{Key Features}}                        \\ \midrule
AMF: Access and Mobility Management Function & Manages access, mobility, and connection states of UEs within the network. & Supports UE authentication, registration, mobility management. \\ \hline
SMF: Session Management Function & Manages session establishment, modification, and termination for data flow between UEs and networks. & Allocates IP addresses, controls QoS, and traffic routing. \\ \hline
UDM: Unified Data Management & Manages user subscription and authentication of data across the network. & Handles user authentication, access control, and service profiles. \\ \hline
NRF: Network Repository Function & Maintains a catalog of available network functions and their capabilities. & Allows for dynamic service discovery and efficient resource allocation. \\ \hline
UDR: User Data Repository & Centralized database for storing structured data, including subscriber profiles and policy information. & Stores subscription data, session data, and policy-related information. \\ \hline
AUSF: Authentication Server Function & Handles authentication for devices and subscribers, ensuring secure access to the network. & Supports advanced authentication methods (\eg, 5G-AKA, EAP-AKA). \\ \hline
PCF: Policy Control Function & Determines and enforces policies for network behavior, including QoS and charging rules. & Centralized policy control, ensures compliance with business logic. \\ \midrule
UPF: User Plane Function & Handles packet routing and forwarding between the UE and external data networks. & Supports low-latency traffic, network slicing, and traffic steering. \\ \bottomrule
\end{tabular}
\vspace{-8pt}
\caption{A list of 5GC network functions and their respective roles.}
\label{tab:5gc-functions}
\vspace{-8pt}
\end{table*}

In the context of 5G, the Mobile Core is transforming into a collection of disaggregated functional components to offer greater flexibility. 
The ETSI document~\cite{3gpptr23.501} outlines the responsibilities of various Network Functions (NFs) and how they communicate with one another. 
Table~\ref{tab:5gc-functions}, based on \cite{3gpptr23.501}, provides a summary of each NF and its main characteristics. 
However, the specific design and implementation of these NFs are left to the discretion of individual vendors.

\begin{figure*}[t]
    \centering
     \begin{subfigure}[b]{0.33\textwidth}
         \centering
         \includegraphics[width=\textwidth]{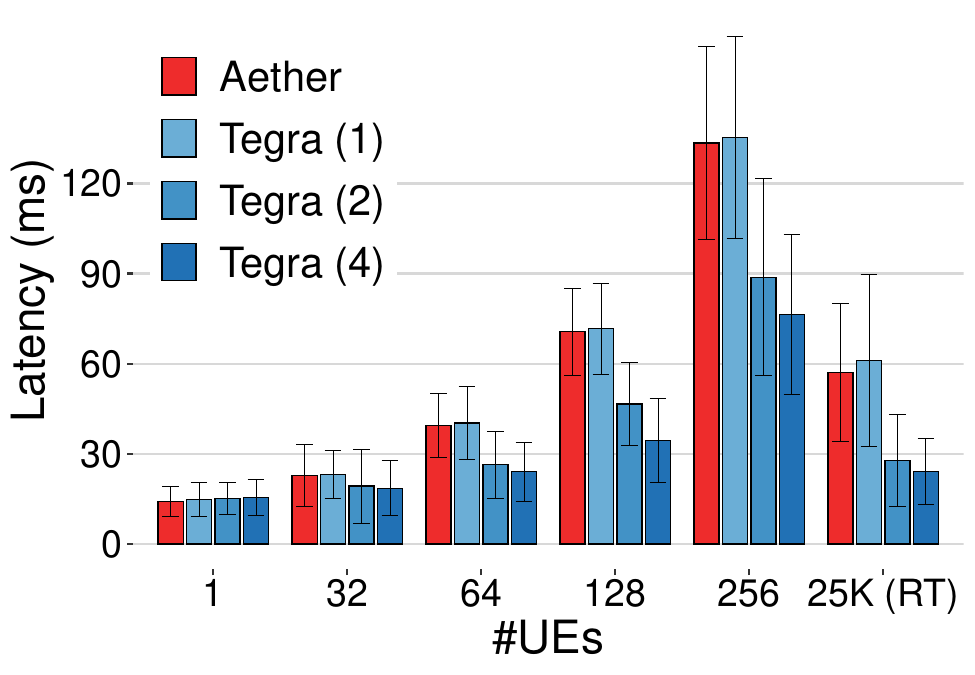}
         \vspace{-0.25in}
         \caption{REG}
         \label{fig:e2e-reg}
     \end{subfigure}
    \begin{subfigure}[b]{0.33\textwidth}
         \centering
         \includegraphics[width=\textwidth]{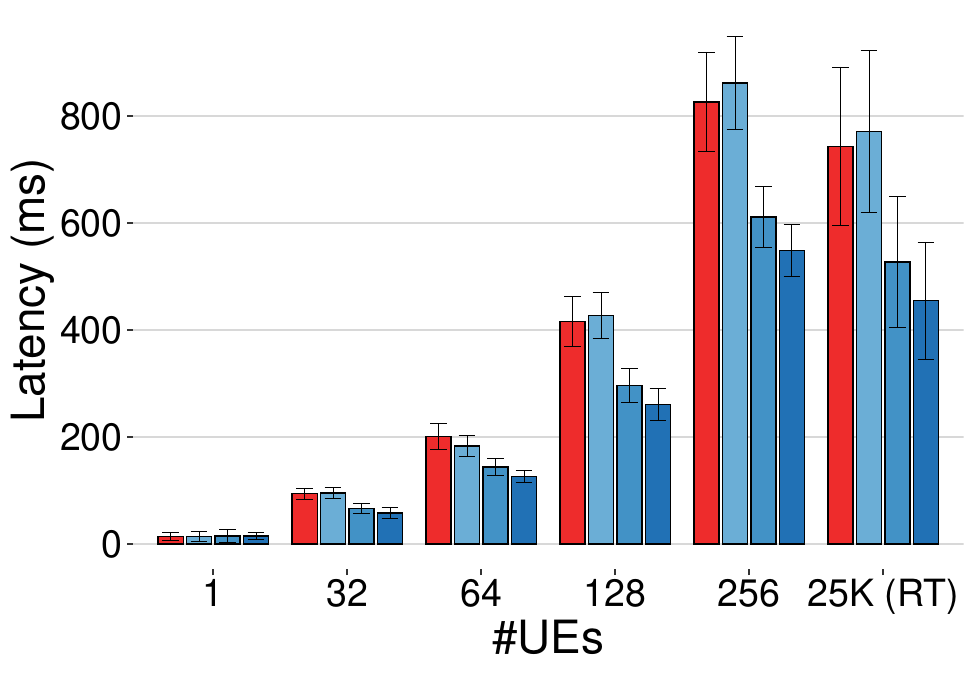}
         \vspace{-0.25in}
         \caption{PDU}
         \label{fig:e2e-pdu}
     \end{subfigure}
     \begin{subfigure}[b]{0.33\textwidth}
         \centering
         \includegraphics[width=\textwidth]{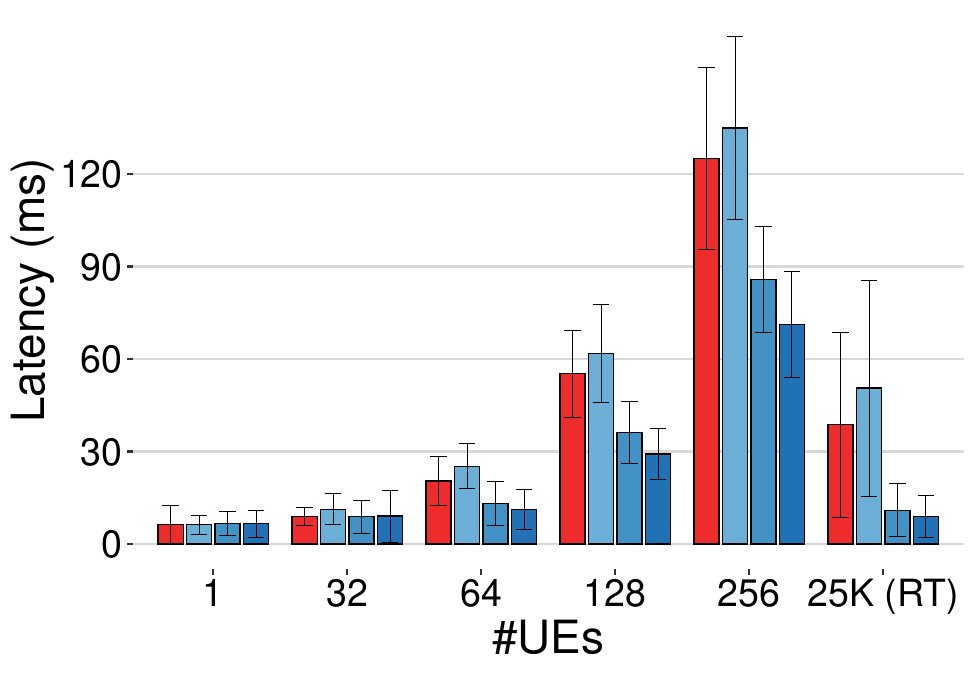}
         \vspace{-0.25in}
         \caption{SR}
         \label{fig:e2e-sr}
     \end{subfigure}
     \vspace{-22pt}
    \caption{Comparison of end-to-end control-plane latency between Aether and \name{} (with the increasing number of 5GC service instances: 1, 2, and 4).}
    \vspace{-10pt}
    \label{fig:e2e-analysis-setup}
\end{figure*}

\section{Using Static Scaling with Aether and \name{}}
\label{sec:static-scaling}

\Cref{fig:e2e-analysis-setup} illustrates the end-to-end latency comparison between Aether and \name{} for three different events (REG, PDU, and SR).
We deploy \name{} in three different configurations, limiting the number of service instances to 1, 2, and 4.
To minimize the warm-up cost, we initiate these instances at the beginning of the run, enabling the requests to be load-balanced among them from the start.
Note that Aether is vertically scaled (each instance can utilize the maximum number of cores if required); however, it lacks the ability to scale by adding more service instances.
On the contrary, \name{} supports horizontal scaling while maximally utilizing the provided cores.

Both Aether and \name{} deliver comparable performance with one UE.
However, as the number of UEs increases, \name{} (with 2 and 4 instances) significantly outperforms Aether, achieving latency reductions of up to 42\% and 33\% with REG and PDU events, respectively, for synthetic traces and 256 UEs.
On real traces with 25K UEs, \name{} delivers 57\% and 38\% reduction for REG and PDU events. 
The most notable improvement occurs with SR events, where \name{} reduces latency by 43\% for synthetic traces and by 76\% for real traces with 256 and 25K UEs, respectively. 
When running with only one instance, \name{} delivers comparable performance to Aether across all scenarios.